\documentclass[twocolumn,tighten]{aastex631}
\usepackage{amsmath}
\usepackage{color}
\usepackage{booktabs}
\usepackage{wrapfig}
\usepackage{appendix}
\usepackage{longtable}

\newcommand{\rproj}{{r_{\rm{proj}}}}
\newcommand{\ruwe}{{\rm{RUWE}}}

\newcommand{\kpc}{{{\rm{kpc}}}}

\newcommand{\Gyr}{{{\rm Gyr}}}
\newcommand{\Myr}{{{\rm{Myr}}}}
\newcommand{\gaia}{{{\em{Gaia}}}}
\newcommand{\ncl}{{N_{\rm{cl}}}}
\newcommand{\nclcg}{{N_{\rm{cl,CG20}}}}
\newcommand{\nclhunt}{{N_{\rm{cl,Hunt}}}}
\newcommand{\pmem}{{P_{\rm{mem}}}}
\newcommand{\pcluster}{{P_{\rm{mem,c}}}}
\newcommand{\pmemcg}{{P_{\rm{mem,CG20}}}}

\newcommand{\pcl}{{P_{\rm{cl}}}}
\newcommand{\pfield}{{P_{\rm{field}}}}
\newcommand{\plx}{{\pi}}
\newcommand{\promo}{{\vec{\mu}}}
\newcommand{\pmra}{{\mu_{\rm{RA}}}}
\newcommand{\pmdec}{{\mu_{\rm{dec}}}}
\newcommand{\fcl}{{f_{\rm{cl}}}}
\newcommand{\pc}{{\rm{pc}}}
\newcommand{\ffield}{{f_{\rm{field}}}}
\newcommand{\emcee}{{\tt emcee}}
\newcommand{\msdata}{{\vec{d}}}
\newcommand{\age}{{t_{\rm{age}}}}
\newcommand{\metal}{{[\rm{Fe/H}]}}
\newcommand{\reddening}{{A_{{v}}}}
\newcommand{\mist}{{\tt MIST}}
\newcommand{\yr}{{\rm{yr}}}
\newcommand{\msun}{{\rm{M}_{\odot}}}
\newcommand{\cgcat}{{CG-DR2}}
\newcommand{\cgmem}{{CG$20$}}
\newcommand{\hunmem}{{HUNT$23$}}

\newcommand{\phasespace}{{\{\promo,\ \plx\}}}

\newcommand{\rdstar}{{r^{\ast}_{\rm{proj}}}}
\newcommand{\sigmacluster}{{\Sigma_{\rm{cluster}}}}
\newcommand{\sigmafield}{{\Sigma^{avg}_{\rm{field}}}}
\newcommand{\nref}{{n_{\rm{ref}}}}
\newcommand{\ra}{{RA}}
\newcommand{\dec}{{dec}}
\shorttitle{Membership and Cluster Properties}
\shortauthors{Ganguly et al.}
\begin{document}
\title{Open Cluster Study Using \gaia\ I:\\Membership and Cluster Properties} 
\author[0000-0002-0991-8438]{Anindya Ganguly}
\affiliation{Tata Institute of Fundamental Research, Department of Astronomy and Astrophysics, Homi Bhabha Road, Navy Nagar, Colaba, Mumbai, 400005, India}
\email{anindyaganguly12@gmail.com}
\author[0000-0002-4638-1035]{Prasanta K. Nayak}
\affiliation{Tata Institute of Fundamental Research, Department of Astronomy and Astrophysics, Homi Bhabha Road, Navy Nagar, Colaba, Mumbai, 400005, India}
\affiliation{Instituto de Astrofísica, Pontificia Universidad Católica de Chile, Av. Vicuña MacKenna 4860, 7820436, Santiago, Chile}
\author[0000-0002-3680-2684]{Sourav Chatterjee}
\affiliation{Tata Institute of Fundamental Research, Department of Astronomy and Astrophysics, Homi Bhabha Road, Navy Nagar, Colaba, Mumbai, 400005, India}
\email{souravchatterjee.tifr@gmail.com}
\begin{abstract}
Star clusters are interesting laboratories to study star formation, single and binary stellar evolution, and stellar dynamics. We have used the exquisite data from \gaia's data release 3 (DR3) to study $21$ relatively rich and nearby open clusters with member numbers ($\ncl$)$>500$. We have developed a non-parametric method to identify cluster members. Our method works well for clusters located in both sparse and crowded environments, hence, can be applied to a wide variety of star clusters. Since the member classification scheme does not make any assumptions on the expected distributions of potential cluster members, our method can identify members associated with clusters that are oddly shaped or have complex internal spatial or kinematic structures. In addition, since the membership determination does not depend on the proximity to any well-defined sequences on the color-magnitude diagram, this method easily identifies straggler members.  
Furthermore, for each of these clusters, we estimate essential cluster properties including age, metallicity, distance, and reddening using detailed Markov-Chain Monte Carlo parameter estimation. We report the full posteriors for these important cluster properties for all clusters in our study.
\end{abstract}
\section{Introduction} \label{sec:intro}
\label{S:intro}
It is believed that stars are born in groups of a variety of sizes, some of which disperse and populate the field, while the others remain gravitationally bound to form star clusters we observe today \citep{LD1991,bra2006,kruij2011,pfalzner2012, Parmentier2013}. Among the most manifest objects in the sky, star clusters are interesting for many branches of astrophysics because they are the sites of the highest density of mostly coeval aggregate of stars with similar metallicity ($\metal$) within a given angular scale. Star clusters have often been used for constraining single and binary stellar evolution \citep[e.g.,][]{Leiner_2021}. The age ($\age$), $\metal$, and the main-sequence (MS) turn-off of a star cluster can be constrained from its colour-magnitude diagram (CMD) with a much greater accuracy compared to individual stellar objects. Because of the clearly defined CMDs, a plethora of exotic stellar sources such as blue-stragglers \citep[BSs, e.g.,][]{Sandage1953,johnson_1955,geller_2011,Geller_2013,Gosnell_2014,brogaard_2018,sindhu_2019,sahu2019,leigh_2019}, sub-subgiant stars \citep[SSGs, e.g.,][]{Geller_2017,leiner_2_2017,geller_3_2017}, and MS-white dwarf (WD) binaries \citep{JADAV_WD_MS_IN_M67_PROCE_2024,grondin_2024}
are easier to identify in star clusters. In addition, dense star clusters are efficient factories of exotic stellar sources formed via various dynamical and binary stellar evolution processes active in them \citep{knigge_2009,Bhattacharya_2019,Nine_2020, Rain_2021, vaidya_2020}. 

The relatively less dense, young, and metal-rich star clusters in the Milky Way (MW) disk are usually called open clusters (OCs). The OCs in the MW exhibit a wide range in $\age$, spanning a few Myrs to several Gyrs \citep[e.g.,][]{Cantat-Gaudin_review_2022}. OCs have been studied for a very long time, some of the first studies date back to the time of naked-eye observations. In the age of modern astronomy, many past surveys have already identified and studied OCs in the MW \citep[e.g.,][]{Dreyer1888}. However, the large and precise astrometric and photometric survey data provided recently by \gaia\ \citep{gaia_dr3} have significantly improved our ability to identify OCs and ascertain membership of objects within them. 

Since OCs are within the MW disk and have much lower stellar densities compared to their more massive and older counterpart, the globular clusters (GCs), it is usually very hard to determine stellar membership for OCs, especially out to a large-enough distance from the OC's fiducial center. In the pre-\gaia\ era, the state-of-the-art was detailed focused observations of specific rich OCs which allowed extraction of stellar populations in the target cluster with great details \citep[e.g.,][]{wiyn1, wiyn2,wiyn27,wiyn28}. Notable exceptions include surveys such as UCAC3 \citep{zacharias2010} and UCAC4 \citep{zacharias2013} which allowed different groups to characterize star clusters and provide detailed catalogs \citep[e.g.,][]{scholz2015,sampedro2017}. With \gaia's precise astrometric and photometric data, the priorities have shifted towards the identification of vast numbers of OCs and characterisation of membership to a limited distance from the fiducial density centre using primarily machine-learning (ML) algorithms \citep[e.g.,][]{Cantat_Gaudin2018a,Cantat_Gaudin2018b,Cantat-Gaudin_upmask2019,ML_aproach1,ML_aproach2,HUNT_COMPARISON,perren_2023}. In particular, thousands of OCs were identified using \gaia's DR2 and several previously identified OC candidates were refuted as asterisms \citep[][hereafter collectively \cgcat]{Cantat_Gaudin2018a,Cantat_Gaudin2018b,Cantat-Gaudin_upmask2019}. Some other studies employed various parameterised statistical techniques such as gaussian mixtures to identify cluster members using proper motion ($\promo$) and parallax ($\plx$) measurements from \gaia's DR2 \citep[e.g.,][]{monterio2019}. Once the OCs are identified and cluster memberships are assessed, it is crucial to characterise the fundamental OC properties including $\age$, $\metal$, the distance to the OC ($D$), and reddening ($\reddening$). Usually, these studies use membership data from previous studies and characterise the OC properties 
\citep[as an example of exception,][assign memberships and find OC properties]{Monteiro2017}. The largest recent study directly adopting the \cgcat\ memberships was done by \citet{Bossini2019} using a Bayesian framework {\tt BASE9} \citep{BASE9_1, BASE-9_2}. However, this work used fixed $\metal$ while determining the OC properties, either at the solar value or adopted from past constraints where available. Using \gaia\ DR2 data, \citet[][hereafter \cgmem]{CG_20} estimated global cluster properties such as $\age$, $\reddening$, and $D$. In a recent study,  \citet[][hereafter \hunmem]{hunt_2023} used Hierarchical Density-Based Spatial Clustering of Applications with Noise (HDBSCAN) to identify clusters in \gaia\ DR3 data up to magnitude $G\sim 20$. Then, they verified these cluster members through a statistical density test and a CMD classifier based on a Bayesian neural network (BNN). In this way, they create one of the largest homogeneous catalogues of OC and their properties. In a different study, \citet{perren_2023} used the \texttt{FASTMP} method to identify members of almost 14,000 star clusters.

In this work, we present the first step of a systematic in-depth study of the rich ($\ncl>500$) nearby ($D/\kpc<3$) OCs. We use \gaia's DR3 to first ascertain membership and then constrain the fundamental properties using a Bayesian framework.  In addition, we aim to reduce subjective human intervention in constraining the fundamental properties using isochrone fitting. Furthermore, we are interested in being able to identify exotic stellar members including BSs and WD binaries that do not fall into any well characterised sequence on the CMD in future work. This motivates us to combine the rigors of the studies on specific OCs \citep[e.g.,][]{wiyn22} and the uniformity of analysis in automated OC surveys using ML methods (e.g., CG-DR2), where it is not easy to interpret the meaning of the membership statistic. Because of the extraordinary success of ML studies to identify OCs using \gaia's DR2, we do not focus on identifying or finding OCs. Instead, here we embark on an in-depth study using \gaia's DR3 \citep[which provides significant improvements in systematics and astrometric solutions][]{gaia_dr3} of rich nearby OCs already mentioned in the \cgmem\ catalog. 

While a vast body of past studies exists on this topic, our study is different from them with several key improvements and flexibility. We first identify the OC members and then with those members, we also estimate the cluters' global properties. We use a new completely non-parametric approach in our analysis, both for membership assessment and estimation of OC properties. Our method automatically takes into account highly correlated error matrices associated with the astrometric and photometric properties of individual stellar objects. Our method is able to handle OCs located in sparse as well as crowded fields and is able to extract membership probabilities even when the phase-space density of field stars starts becoming comparable (or even higher) than that of cluster stars. Once the members are identified, we estimate OC fundamental properties using a Bayesian framework with no subjective intervention. This is traditionally challenging because the MS on a CMD can show a spread due to photometric errors, $\reddening$, and existence of stellar binaries. Thus, identifying the single MS and fitting an isochrone is usually not trivial without human supervision. Nevertheless, recent advancements do allow hands-off approaches to fit isochrones taking into account the presence of stellar binaries \citep[e.g.,][]{Perren_2015,Monteiro2017,monterio2019}. We introduce a different approach which is relatively less computationally expensive and allows us to make no assumptions for the stellar mass function or the stellar binaries. 

The rest of the paper is organised as follows. In \autoref{S:data_selection} we describe the details of how we select the data. In \autoref{S:membership} we describe how we calculate the membership probabilities. In \autoref{S:cluster_props} we describe how we constrain the fundamental properties of the OCs using the member stars. In \autoref{S:results} we show our key results and conclude in \autoref{S:conclusion}. 

\section{Data Selection} \label{S:data_selection}
We select 21 OCs from the \cgmem\ catalog that are not within $\pm5^\circ$ of the Galactic plane, has $\ncl>500$ (as mentioned in \cgmem), and are within a distance $D/\kpc<3$. Although the methods we have developed can be applied to crowded as well as sparse fields, as a first application, we have chosen to be sufficiently away from the Galactic plane. While we select the target OCs based on \gaia's DR2 used by \cgmem, we use DR3 data for our analysis. In DR3, on an average, the astrometric precision in $\plx$ ($\promo$) has increased by $\sim30\%$ ($\sim$ factor of 2). The systematic errors in astrometry are suppressed by $30$--$40\%$ for $\plx$ and by a factor of $\sim2.5$ for $\promo$. DR3 also provides better precision in photometry with no systematics above $\sim1\%$ in any bands $G$, $BP$, and $RP$ \citep[e.g.,][]{gaia2016b, gaia_dr3}. These improvements are expected to help ascertaining cluster membership and finding the cluster's global properties. 

For each of our selected OCs, we analyze {\em all} \gaia\ sources with $G<18$ within a projected distance (from the fiducial cluster center) $\rproj=20\,\pc$. Neither of these cuts are explicitly required in our analysis and introduced to limit computational cost. As an example, increasing the magnitude by 1 introduces a $\approx45\%$ increase in the number of sources we need to analyze. We also noticed that fainter stars exhibit larger astrometric and photometric errors and we find that including them in the analysis does not change the estimated cluster properties at all. The cut on $\rproj$ is essentially to limit the total number of \gaia\ sources we need to analyze, hence, limiting computational cost. At the same time, we do not want to limit the analysis only to the central regions of the clusters. Our adopted $\rproj/\pc=20$ is $\gtrsim3$ times the radius containing $50\%$ of the cluster members, $r_{50}$, reported in \cgmem\ for all OCs we study. Moreover, we find that in all our analyzed OCs, the projected number densities of field stars become significantly higher than that of the OC members well below $\rproj/\pc=20$ (see \autoref{S:surface_density_profile} for more details). This gives us belief that extending our analysis to significantly larger $\rproj$ may not yield a significant increase in cluster members. In any case, identifying the true spatial spread of star clusters is not the goal of this study. 

In contrast to several earlier studies, we do not impose any cuts based on the Renormalised Unit Weight Error (RUWE) for the sources. While, $\ruwe>1.4$ may indicate an inadequate single star astrometric solution for a source, past studies indicate that the high $\ruwe$ is often obtained for stellar binaries \citep{jorissen2019impact,binary_ruwe}. Since, imposing an upper limit on $\ruwe$ may preferentially exclude stellar binaries, we choose not to impose a hard $\ruwe$ cut-off. Nevertheless, we take into account $\plx$ and $\promo$ errors for each source by using the three-dimensional Gaussian's where the mean is given by the fiducial values for the source and the covariance matrix is constructed from the reported $1\sigma$ errors in \gaia-DR3. This allows us to, for example, put more weight on stars that have small errors in $\promo$ and $\plx$. 
\section{Membership analysis}
\label{S:membership}
The OCs being part of the disk population of the MW, with low central stellar densities, clearly identifying cluster members through a combination of RA, Dec, $\plx$, and $\promo$ poses challenges. The collection of stars to consider and contribution from the field increases with increasing $\rproj$. A small enough $\rproj$ may allow identification of cluster stars with $\promo$ significantly different from those in the field. However, in the crowded fields of the typical OCs, increasing $\rproj$ may lead to such dominance of field stars that it becomes impossible to easily separate the cluster from the field even after considering $\plx$ and $\promo$. This difficulty usually limits the ability to consider stars with sufficiently large $\rproj$. 

We develop a non-parametric method that can extend to arbitrarily large $\rproj$, the limitation coming only from the fact that the number of stars to analyze increases significantly with increasing $\rproj$. We calculate the cluster membership probability 
\begin{equation}
    \pmem = \frac{\pcl}{\pcl+\pfield},
    \label{eq:pmember}
\end{equation}
where $\pcl$ ($\pfield$) denotes the probability that a particular star belongs to the cluster (field), calculated based on the three dimensional parameter space $\plx$, $\promo$ and the associated errors. 
\subsection{Construction of the cluster's probability density function}
\label{S:cluster_pdf}
In order to evaluate $\pcl$ for any particular star with a given $\{\pmra,\ \pmdec,\ \plx\}$, we first need to construct a three-dimensional probability density function (PDF) for cluster stars, $\fcl(\pmra,\ \pmdec,\ \plx)$. This presents a chicken-and-egg problem, since identifying cluster members requires some prior knowledge of the cluster itself. Thus, we are forced to take an initial guess of a collection of stars that, with very high probability, are cluster members with little chance of contamination from the field. Selecting the stars we can use as the initial guess of cluster stars is the only step in our analysis which requires human intervention.\footnote{We have plans to improve this in future.} 

Based on several trials, we have decided to analyze all stars with $\rproj/\pc\leq1$ to identify our initial guess of cluster stars by clustering of sources in $\{\pmra,\ \pmdec,\ \plx\}$. \autoref{fig:initial_cluster_guess} illustrates our strategy using a pair-wise three-step process for NGC 2287 as an example. First, we identify the clustered sources in $\{\pmra,\ \pmdec\}$ (top panel, green circles). However, in $\{\pmra,\ \plx\}$, 4 out of these selected sources do not seem to be tightly clustered with the others. Hence, we discard these 4 sources from our selection and consider only the others (middle panel, blue plus). All sources clustered both in $\{\pmra,\ \pmdec\}$ and $\{\pmra,\ \plx\}$ are then found to be clustered also in $\{\pmdec,\ \plx\}$ (bottom panel, red dots) in this particular example. Thus, the red dots constitute the sources with $\rproj/\pc<1$ exhibiting tight clustering separate from the other stars in $\{\pmra,\ \pmdec,\ \plx\}$. These sources constitute our initial guess of cluster stars for this OC. Note that these sources are simply our initial guess for cluster stars and it is important to be very selective because these stars would constitute $\fcl$. A cluster identification algorithm could remove the necessity for human intervention at this stage. However, we find that by eye we are able to be more selective in our choice, especially for OCs in relatively more crowded regions. Our tests suggest that including or excluding a small number of marginal sources does not influence the final $\pmem$ of the much larger number of sources we ultimately analyse. 

\noprint{\figsetstart}
\noprint{\figsetnum{1}}
\noprint{\figsettitle{Illustrations of pair-wise three-step process for initial guess of cluster member sources.}}

\figsetgrpstart
\figsetgrpnum{1.11}
\figsetgrptitle{NGC 2287}
\figsetplot{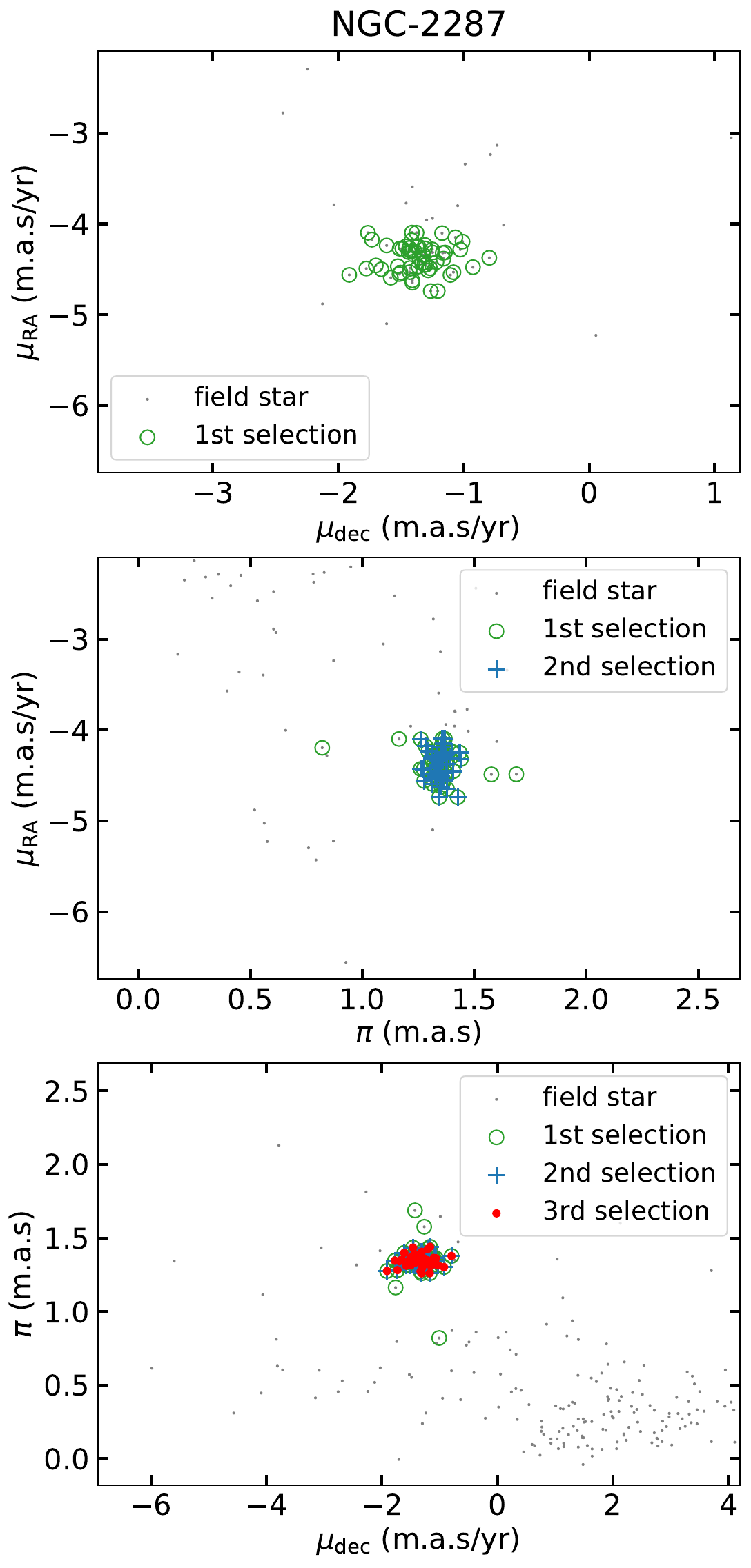}
\figsetgrpnote{Illustration of how we selected the initial guess for cluster member sources of NGC 2287.}
\figsetgrpend

\figsetend

\begin{figure} 
\gridline{\fig{initial_cut_NGC_2287.pdf}{0.40\textwidth}{}}
\caption{Illustration of how we select the initial guess for cluster member sources using NGC 2287. We first select all sources with $\rproj/\pc\leq1$. We identify all sources closely clustered in all of the 
$\{\pmra,\pmdec\}$ (top, green circles), $\{\pmra,\plx\}$ (middle, blue plus), and $\{\pmdec,\plx\}$ planes and discard those that are not closely clustered in any of these planes. For example, in the middle panel we discard four sources (green open circles) and only select the others (blue `+' with green open circle). In the bottom panel we find that the sources closely clustered both in $\{\pmra,\pmdec\}$ and $\{\plx,\pmra\}$, are also closely clustered in $\{\pmdec,\plx\}$ (green circles with blue `+' and red dot). Sources depicted by the green circles, blue `+', as well as red dot constitute our initial guess for cluster members. Grey dots denote other stars that show no clustering in any of the planes.}
\label{fig:initial_cluster_guess}
\end{figure}
Once our initial guess of cluster stars is selected, we proceed to construct $\fcl$. Instead of using the fiducial $\phasespace$ values for the stars, we assume that the $i$th star is represented by a three-dimensional Gaussian $\mathcal{N}_i$ with the mean given by the fiducial values and covariance matrix given by the correlated errors reported in \gaia's DR3.\footnote{If the detailed posteriors for each star were available, we could have used those. In the absence of this, we have to adopt the normal distribution.} The cluster PDF is then $\fcl=\sum_i \mathcal{N}_i/N$, where $N$ is the total number of sources in the initial guess. In practice, we randomly draw from $\mathcal{N}_i$ $100$ times for the $i$th star and combine all draws for all stars to construct $\fcl$ using kernel density estimation (KDE) with Gaussian kernels from the Scipy package \citep{2020SciPy}, bandwidth evaluated using Scott's method \citep{scott_method_book}. Thus, if the number of stars in our initial guess for cluster stars is $N$, $\fcl$ is constructed with $100N$ points. \autoref{fig:construction_of_cluster_kde} shows $\fcl$ for our example cluster NGC 2287 along with the initial guess of cluster stars. 

\noprint{\figsetstart}
\noprint{\figsetnum{2}}
\noprint{\figsettitle{PDFs for sources in our initial guesses of cluster members.}}

\figsetgrpstart
\figsetgrpnum{2.11}
\figsetgrptitle{NGC 2287}
\figsetplot{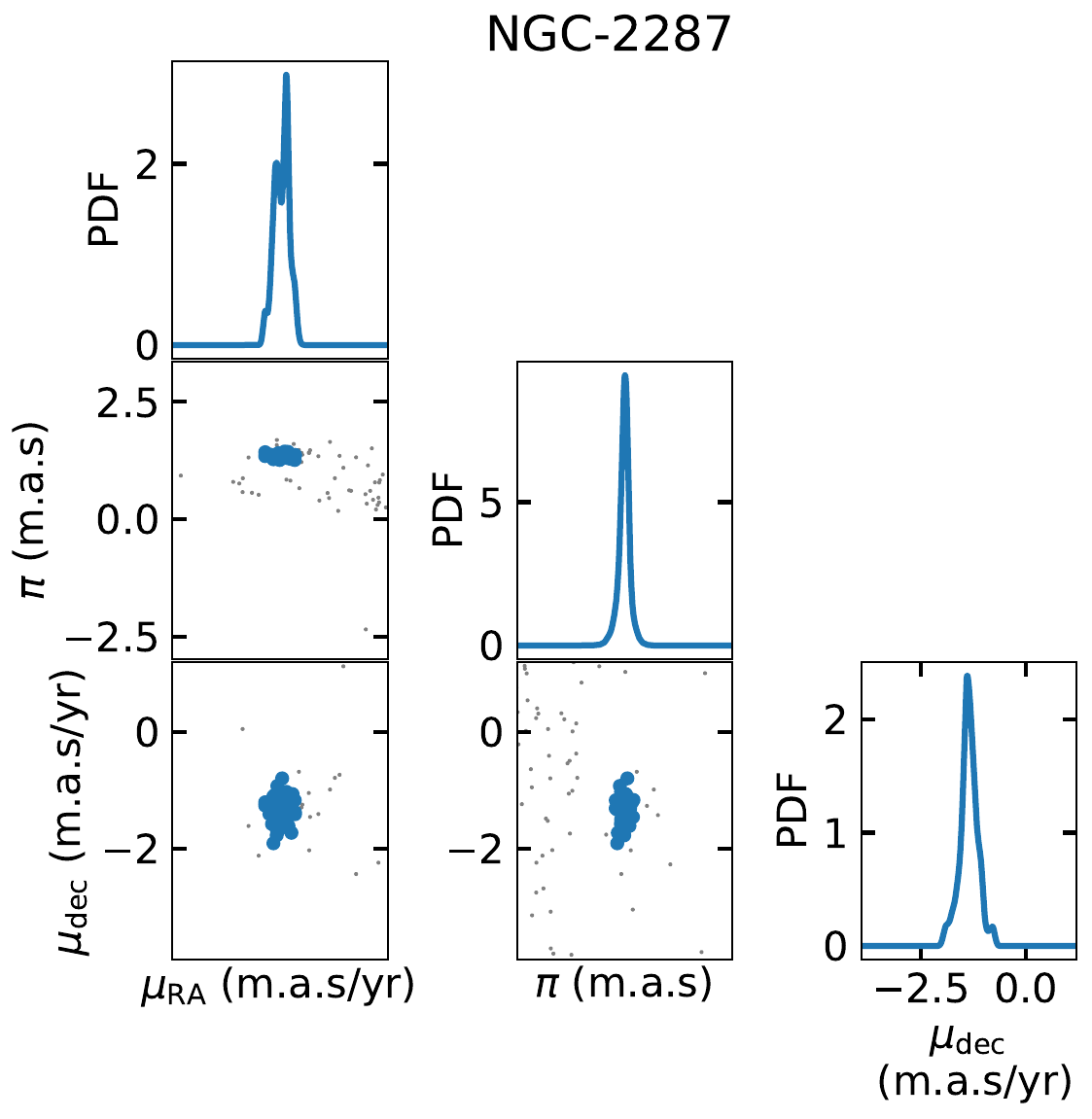}
\figsetgrpnote{PDFs for sources in our initial guess of cluster members of NGC 2287.}
\figsetgrpend

\figsetend

\begin{figure} 
\gridline{\fig{cluster_kde_initial_guess_NGC_2287.pdf}{0.45\textwidth}{}}
\caption{
PDFs for the sources in our initial guess of cluster members (blue dots). Grey dots denote all other sources with $\rproj/\pc\leq1$.
}
\label{fig:construction_of_cluster_kde}
\end{figure}
\subsection{Construction of the field PDF}
\label{S:field_pdf}

\noprint{\figsetstart}
\noprint{\figsetnum{3}}
\noprint{\figsettitle{Normalized histograms and PDFs for sources within a projected distance of 20 pc from cluster centers.}}

\figsetend

\begin{figure} 
\gridline{\fig{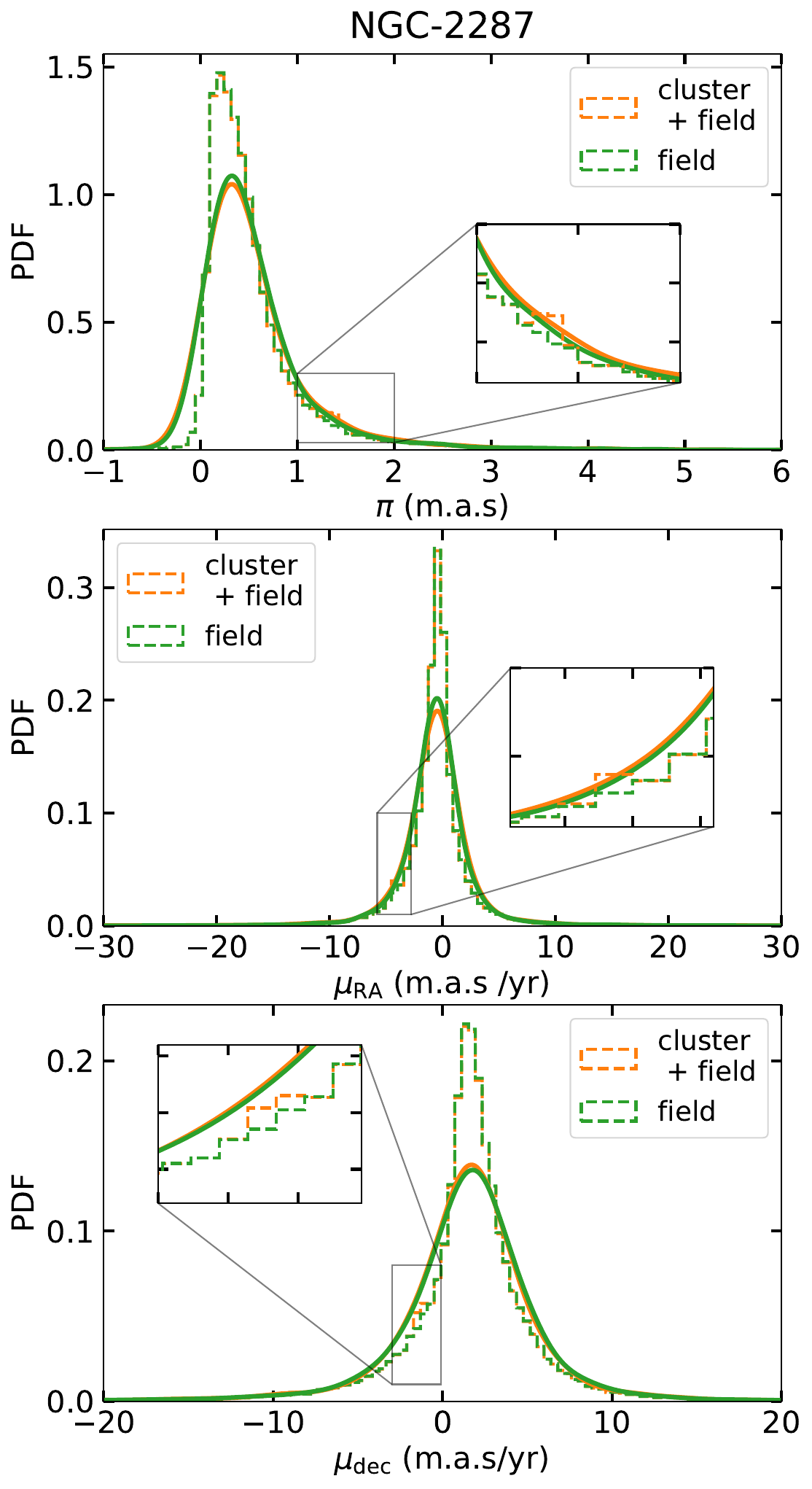}{0.45\textwidth}{}}
\caption{Normalized histogram (dashed) and PDF (solid) for sources within $\rproj/\pc\leq20$ from the cluster center of NGC-2287. Green and orange denote all sources and those excluding our initial guess of cluster members (detailed discussion in \autoref{S:membership} and in \autoref{fig:initial_cluster_guess}). Top, middle, and bottom panels are for $\plx$, $\pmra$, and $\pmdec$, respectively.}
\label{fig:construction_of_field_kde}
\end{figure}
Similar to the construction of $\fcl$, we construct $\ffield$ which gives a reasonable description of the field stars in $\phasespace$. For this, we consider all stars within $\rproj/\pc\leq20$ except those that are very likely cluster members, i.e., stars with $\phasespace$ within the volume containing $95\%$ of the cluster PDF, $\fcl$. 
Using these stars, we construct $\ffield$ the same way as we construct $\fcl$. 
As an example, \autoref{fig:construction_of_field_kde} shows the histogram as well as the PDF for all stars with $\rproj/\pc\leq20$ (orange) for NGC-2287. It also shows the subset of stars except those within the volume containing $95\%$ of total probability of $\fcl$ (green). The tiny over-densities (visible only in the zoomed-in insets) is the contribution of the star cluster. \autoref{fig:construction_of_field_kde} illustrates a serious challenge in identifying OC members. Even for our example OC NGC-2287, which is relatively rich with stars, and even after considering all of the three relevant parameters, $\pmra,\ \pmdec,\ \plx$, the contribution from the OC is rather small compared to those in the field if a reasonably large $\rproj/\pc\leq20$ is considered. This challenge often limits the $\rproj$ up to which stars can be analyzed. In our strategy, we are not limited by this challenge, since after $\fcl$ and $\ffield$ are constructed, we can easily evaluate the membership probability of any star at any $\rproj$. For us, the limit comes from practical considerations and computational cost since the number of stars to analyse increases linearly with $\rproj$.   
\subsection{Cluster Membership probabilities}
\label{S:final_membership}
Once $\fcl$ and $\ffield$ are constructed, it is straightforward to calculate the probability of drawing a particular star from either the cluster or the field PDF. We take into account errors in $\{\pmra,\ \pmdec,\ \plx\}$ measurements for each star in the same way as described above. We draw 100 points from a 3D gaussian created using the fiducial values as mean and the errors as $1\sigma$ covariance. For each of these values, we calculate the probability densities corresponding to the PDF $\fcl$ ($\ffield$) and take the average to finally arrive at the probability $\pcl$ ($\pfield$) of drawing this particular star from $\fcl$ ($\ffield$). Incorporating the astrometric errors instead of simply using the fiducial values allows us to directly factor in accuracy of astrometric measurements of each star into the final uncertainties in determining $\pcl$.\footnote{For example, consider stars 1 and 2, both with fiducial values of $\vec{\mu}$ and $\plx$ close to that of the cluster's, but the former has larger errors than the latter. Using our strategy, $\pcl$ for the former will automatically be higher than that of the latter. In contrast, among two stars, both with fiducial values of $\{\vec{\mu}, \plx\}$ far away from the peak of $\fcl$, the one with larger errors will likely have a higher $\pcl$ because the long tails will allow higher overlap with $\fcl$. Fainter stars typically have larger astrometric errors. Incorporating correlated astrometric errors directly into the analysis also takes that into account.}      

\noprint{\figsetstart}
\noprint{\figsetnum{4}}
\noprint{\figsettitle{Source counts as a function of membership probability.}}

\figsetgrpstart
\figsetgrpnum{4.11}
\figsetgrptitle{NGC 2287}
\figsetplot{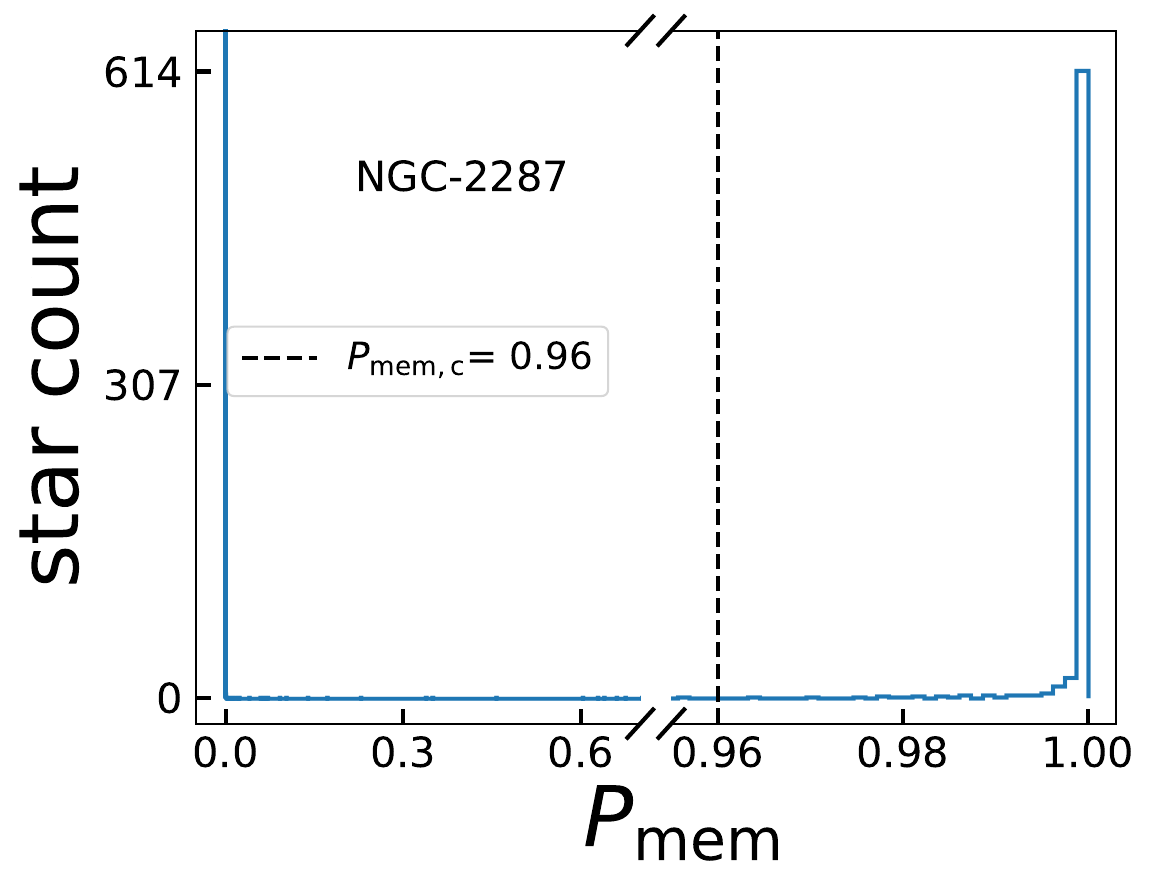}
\figsetgrpnote{Source count as a function of membership probability for NGC 2287.}
\figsetgrpend

\figsetend

\begin{figure} 
\gridline{\fig{pmem_hist_NGC-2287.pdf}{0.47\textwidth}{}}
\caption{
Source count as a function of membership probability $\pmem$. Sources with very low $\pmem$ clearly belong to the field. Source count increases sharply near $\pmem\approx1$. The dashed black vertical line denotes the cut-off value $\pcluster$ in $\pmem$ for this cluster. Very few sources have $P_{\rm{mem}}<\pcluster$ until $P_{\rm{mem}}\sim0.0$. Thus, our evaluated $\pmem$ divides the sources in two clearly distinct groups, cluster members with $\pmem\approx1$ and non-members with $\pmem\approx0$. 
}
\label{fig:memberhip_vs_num_hist_2287}
\end{figure}
We evaluate $\pcl$ and $\pfield$ for all stars with $\rproj/\pc\leq20$ for each OC and calculate the membership probability ($\pmem$) using \autoref{eq:pmember}. \autoref{fig:memberhip_vs_num_hist_2287} shows star count as a function of $\pmem$ for our example OC NGC-2287. The distribution shows two clear peaks populated by high-probability cluster stars and those that clearly do not belong to the cluster. There is a low floor in star count in between the two extreme $\pmem$ values corresponding to the cluster members and non-members. In order to find the optimal cut-off value in the membership probability, $\pcluster$, we first estimate the average level of the flat part, $\nref=\Delta N/\Delta \pmem$, where, $\Delta N$ is the total number of stars within $\Delta \pmem=(0.8-0.5)$. We define $\pcluster$ as the $\pmem$ value above which $\delta N/\delta\pmem>5\nref$ and increases monotonically, where $\delta$ denotes bin-wise values. We consider all stars with $\pmem>\pcluster$ as cluster members.\footnote{The exact cut-off, $\pcluster$, is a matter of choice. One may choose to be more or less selective. We make public $\pmem$ for all stars to enable users to freely choose a different cut-off if desired.
}    

\noprint{\figsetstart}
\noprint{\figsetnum{5}}
\noprint{\figsettitle{All sources within a projected distance of 20 pc.}}

\figsetgrpstart
\figsetgrpnum{5.11}
\figsetgrptitle{NGC 2287}
\figsetplot{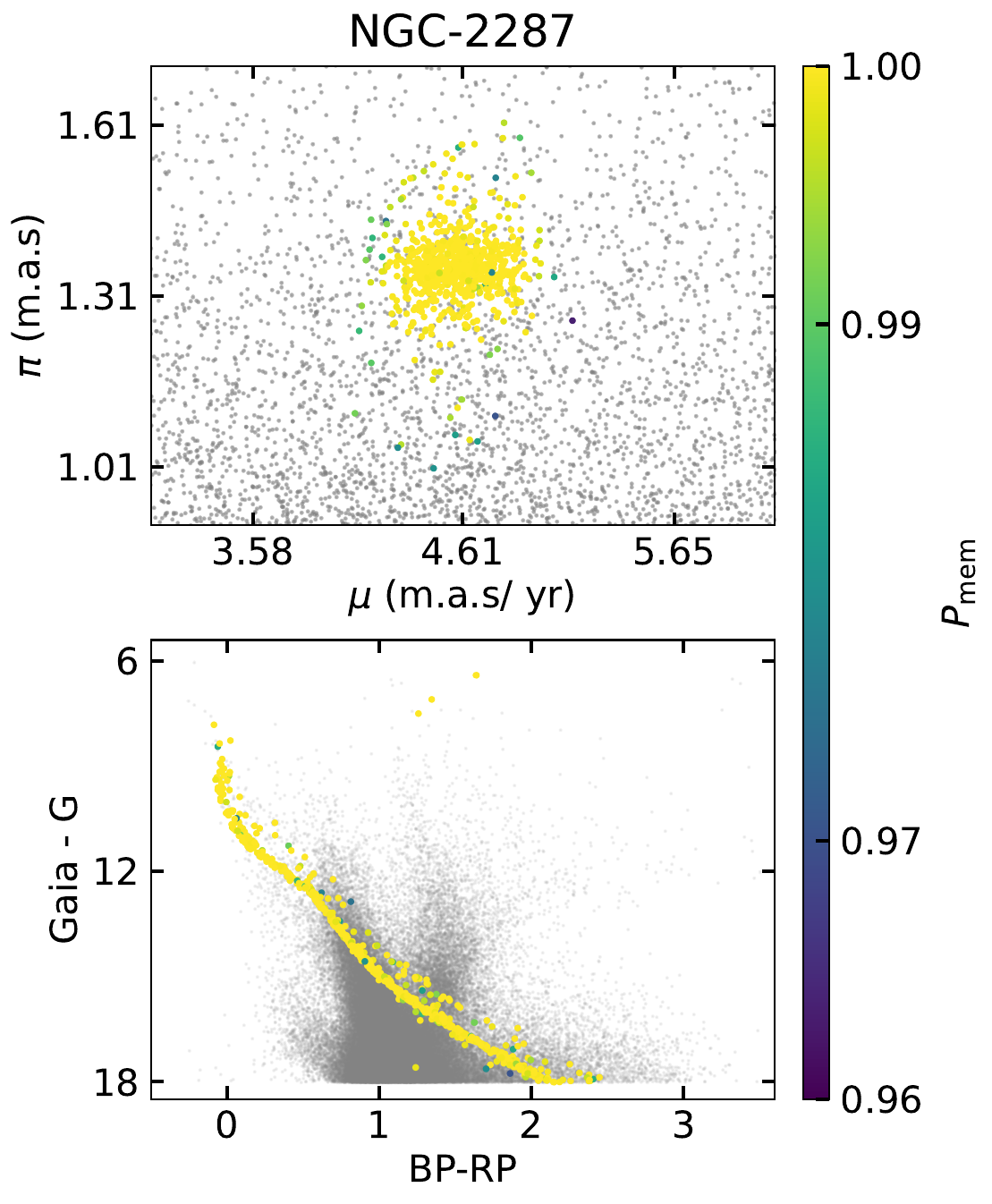}
\figsetgrpnote{All sources within a projected distance of 20 pc for NGC 2287.}
\figsetgrpend

\figsetend

\begin{figure} 
\gridline{\fig{NGC-2287_cluster_membership_5times.pdf}{0.45\textwidth}{}}
\caption{
All sources with $\rproj/\pc\leq20$ for NGC 2287. Colored dots denote sources with $\pmem>\pcluster$, i.e., sources we identify as cluster members, where the color denotes $\pmem$. Sources denoted by grey dots have $\pmem<\pcluster$. 
\textit{Top}: $\plx$ and $|\promo|$ of all cluster members (colored dots) and non-members (grey dots). \textit{Bottom}: CMD for all sources. Clearly, the cluster members produce a very clean CMD. We find a few giant star with very high $\pmem$. Also, as expected, fainter stars have typically lower $\pmem$.
}
\label{fig:cluster_membership1}
\end{figure}
In \autoref{fig:cluster_membership1} we show the CMD (bottom) and $\promo$ vs $\plx$ (top) for all stars with $\rproj/\pc\leq20$ of NGC-2287. The grey dots denote all stars we analyse, the coloured dots denote cluster members based on $\pcluster$ described above. The colour bar denotes $\pmem$. Clearly, stars near the cluster center in $\phasespace$ have very high $\pmem$, while $\pmem$ gradually decreases towards the outskirts. Also note that $\pmem$ decreases faster towards the $\phasespace$ with a higher field star density due to higher contamination. These are all expected and desirable behaviours. 

\noprint{\figsetstart}
\noprint{\figsetnum{6}}
\noprint{\figsettitle{PDFs and scatter plots for initial guesses and final identified cluster members.}}

\figsetgrpstart
\figsetgrpnum{6.11}
\figsetgrptitle{NGC 2287}
\figsetplot{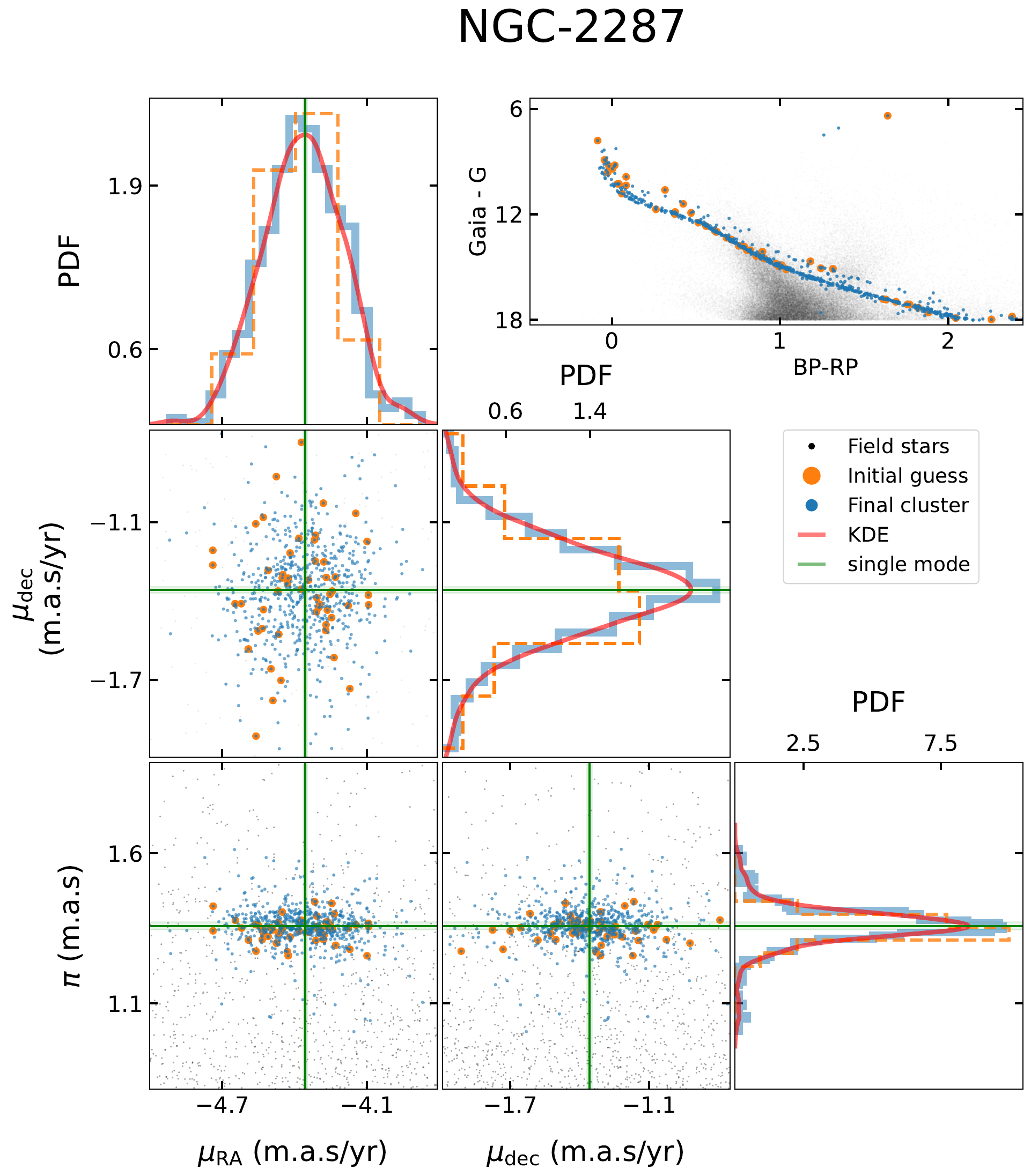}
\figsetgrpnote{PDFs and scatter plots for our initial guess and final identified cluster members for NGC 2287.}
\figsetgrpend

\figsetend

\begin{figure*}
\gridline{\fig{initial_final_cluster_NGC-2287_v_mode.pdf}{0.9\textwidth}{}}
\caption{The PDFs and scatter plots for our initial guess (orange) and final identified (blue) cluster members in $\{\promo, \plx\}$ for NGC-2287. In the top right corner, we show the CMD. Grey dots denote non-members. The solid red line shows the KDE for the distributions created using a gaussian kernel with a bandwidth adopted using the Scott's method. The green (blue) line marks the position of the primary mode (secondary mode, where available).
}
\label{fig:2287_corner_ini_final_plot}
\end{figure*}
In \autoref{fig:2287_corner_ini_final_plot}, we compare $\pmra$, $\pmdec$, and $\plx$ for our initial guess (orange) and final (blue) cluster members for NGC-2287 as an example. The histograms show the one-dimensional distributions and dots denote the sources. Grey dots denote sources that are not members of NGC-2287. We also show the CMD and show the relative positions of our initial guess and final clusters members. The excellent alignment in parameter space between the initial guess and final cluster members illustrates the validity of our initial guess and the overall strategy to identify cluster members. Also note that the sources that constitute our initial guess for $\fcl$, span the entire range of the CMD indicating that our choice of sources were not biased. 

\subsection{Example clusters at different difficulty levels}
\label{S:example_cluster}
\begin{figure*} 
\gridline{\fig{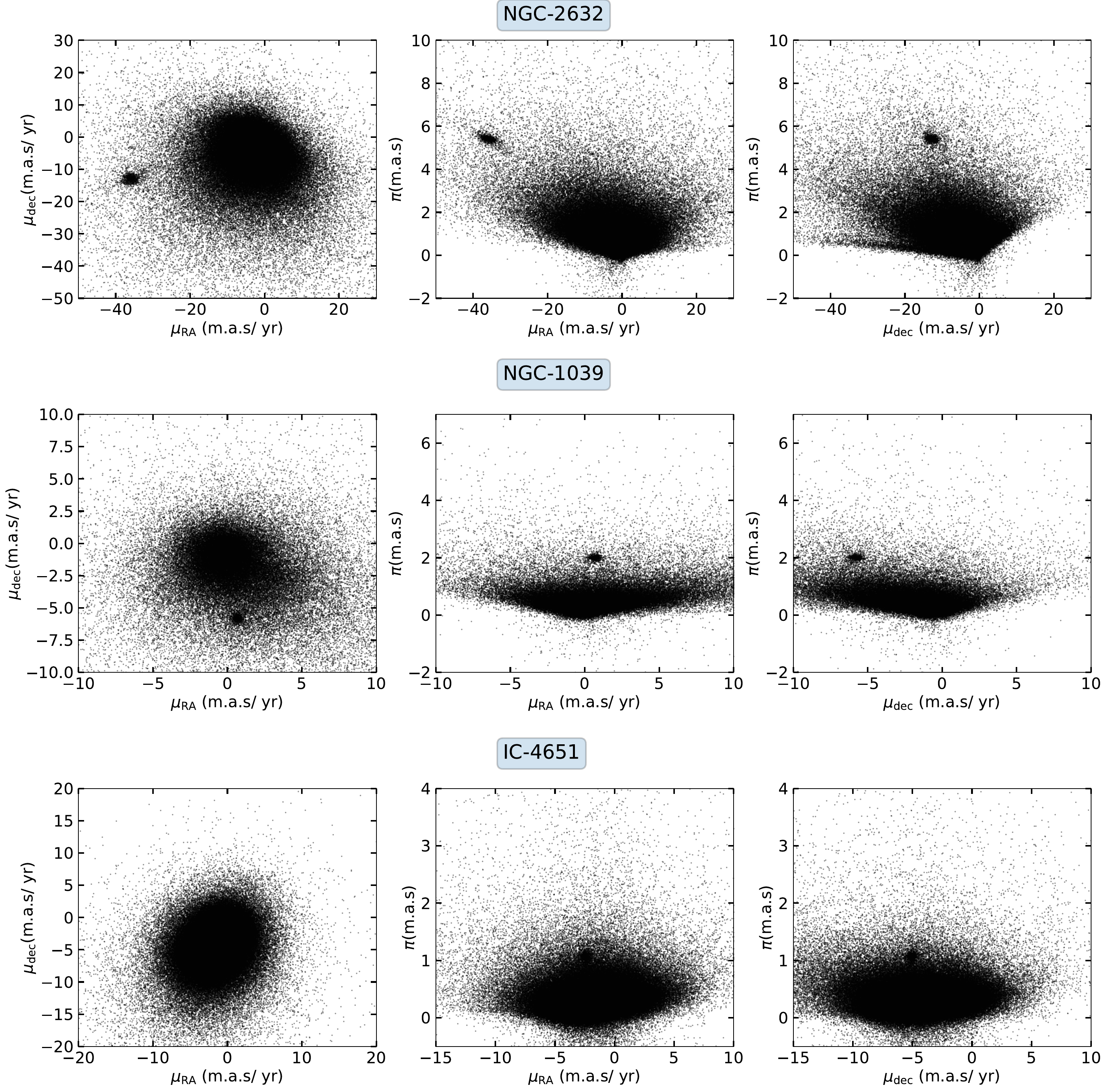}{0.95\textwidth}{}}
\caption{Two dimensional sections of $\pmra$, $\pmdec$, and $\plx$ of all sources (dots) within $\rproj/\pc=20$ for example clusters representing three different levels of overlap with the field stars and hence difficulty in identification. The cluster NGC-2632 (top) is well separated from the field stars and can be easily identified in all astrometric planes. For NGC-1039 (middle), the cluster has a higher overlap with the field stars. It is still identifiable in the planes involving $\plx$, but not easily identifiable without it. 
IC-4651 (bottom) has a very high degree of overlap with the field and it is not easy to identify separately from the field stars in any of the astrometric planes. 
}
\label{fig:easy_medium_hard_cluster}
\end{figure*}

While we illustrate our algorithm to identify cluster members using NGC 2287, a rich evolved cluster, as an example, our sample of clusters contain clusters at widely differing levels of difficulty primarily determined by the degree of their overlap in parameter space with nearby field stars. 
\autoref{fig:easy_medium_hard_cluster} shows all sources within $\rproj/\pc=20$ for three representative clusters with different levels of difficulty in identifying the cluster members. For each source we use the fiducial values and ignore the errors in the astrometric measurements. In the case of NGC-2632 (top), we can easily identify the over-density of the whole cluster in all three astrometric planes. Clearly, identifying the cluster members is relatively easy in this case. For NGC-1039 (middle) the over density corresponding to the cluster is less clear. Nevertheless, it is still possible to identify the cluster in the $\{\pmra,\pi\}$ and $\{\pmdec,\pi\}$ planes. It is however, barely distinguishable in the $\{\pmra,\pmdec\}$ plane. In the case of IC-4651 (bottom), the situation is even worse. The cluster is not distinguishable in any of the astrometric planes. Note that the overlap between the cluster and field stars appears even worse if the associated uncertainties in the astrometric properties for each source is considered. Clearly, it is not easy to identify the cluster members with such a large adopted $\rproj$. Our goal for this study is to develop a procedure that can be applied without any loss of generality or special attention to clusters with any degree of overlap with the nearby field stars. Using a very restrictive limit of $\rproj/\pc=1$ to identify the sources that are most likely cluster members allows us to achieve this. Once the astrometric parameter space associated to the cluster members, $\fcl$, is clearly identified using sources only within $\rproj/\pc=1$, sources within an arbitrarily large $\rproj$ can be considered. In all of our sample clusters with dramatically different levels of difficulties and overlap with field stars, this procedure can be applied without any loss of generality.     

\section{Estimation of cluster properties}
\label{S:cluster_props}
Once the cluster members are selected, we can estimate the clusters' global properties. The distributions of astrometeric parameters are, in general, unimodal. Hence, we have mentioned the median, $16$th, and $84$th percentile value of astrometric properties, $\pmra$, $\pmdec$, and $\plx$ of the clusters in \autoref{tab:cluster_properties_table}. However, for Collinder-69 and Melotte-20, we do notice a bimodal distribution in $\pmra$. To investigate this in detail, we create KDEs of all the astrometric properties for 21 clusters using gaussian kernels and Scotts's bandwidth method \citep{scott_method_book}, and estimate the mode values from the KDEs, which clearly show the presence of two modes in the $\pmra$ distribution of Collinder-69 and Melotte-20 (please see the figure set of \autoref{fig:2287_corner_ini_final_plot}). In \autoref{fig:2287_corner_ini_final_plot}, the red thick solid lines show the KDE for NGC-2287 for $\pmra$, $\pmdec$, and $\plx$ and the green lines depict the mode (in this case single) value of these distributions.

Using the cluster members we identify, we estimate cluster properties such as $\age$, $\metal$, $D$, and $\reddening$. 
Our method uses a robust Bayesian framework to estimate the posterior distributions of these parameters using isochrone fitting and requires no visual inspection. As a result, the same technique can be uniformly used for all clusters. Below we discuss the detailed procedure. 
\subsection{Identification of the single-star main sequence}
\label{S:cluster_props_MS}
The primary challenge in fitting an isochrone to a CMD without any visual inspection is in the identification of the single-star MS. The presence of unresolved binaries as well as photometric errors, which also often depends on the magnitude, create a spread in the CMD of a star cluster. While it is straightforward to visually understand where the single-star main sequence is on the CMD, it isn't so without visual inspection. Below we explain the technique we have developed that can be used self-consistently for any cluster. 

It is expected that the total binary fraction in an OC is $\lesssim50\%$ for GKM stars \citep[e.g.,][]{jadhav2021}. This means that for any given range in magnitude, the highest density of sources on the CMD should correspond to the single-star main sequence. On the other hand, while the individual photometric errors can be large, they are expected to be random and centred around the actual value. Hence, while photometric errors can spread the single-star MS, if the colors and magnitudes for each source are treated as a distribution considering the fiducial values and associated errors, the source density is expected to remain roughly near the correct location provided there are enough sources to consider. 

We exploit the above expectations in the following way. We first sort the MS stars in G. Then from the top of the MS, we take a sliding window of 30 sources and create a two-dimensional PDF in G and BP-RP using the reported fiducial values and associated errors in \gaia's DR3 assuming Gaussian distributions for each source.\footnote{We have tested by varying the size of this sliding window. Choosing a large number reduces resolution at the top of the MS near the turn-off region. On the other hand, choosing too small a number increases fluctuations significantly. We find that the adopted size provides the most optimal outcome for the OCs we have considered.} We find the location of the dominant mode of the PDF for each group. In rare occasions, if we find two modes of roughly equal importance, we choose the mode at the lower BP-RP since the other mode is expected to come from stellar binaries. For all sources, we find that DR3's errors in BP-RP are significantly higher compared to that in G. Hence, we adopt the $99$th percentile value in the errors in BP-RP for all sources in the corresponding group as a conservative estimate for the error in the location of the mode. At the end of this exercise, we obtain the mode locations and associated errors for all sources within small slices in G across the whole CMD of the OC. 

We fit a degree-$5$ polynomial through the mode locations on the CMD. We use this best-fit polynomial to define the locus of the single-star MS on the CMD. Using the associated errors in the locations of the modes, we fit two separate degree-$5$ polynomials depicting the spread below and above the single-star MS.  

\noprint{\figsetstart}
\noprint{\figsetnum{8}}
\noprint{\figsettitle{Distance distributions of identified cluster members.}}

\figsetgrpstart
\figsetgrpnum{8.11}
\figsetgrptitle{NGC 2287}
\figsetplot{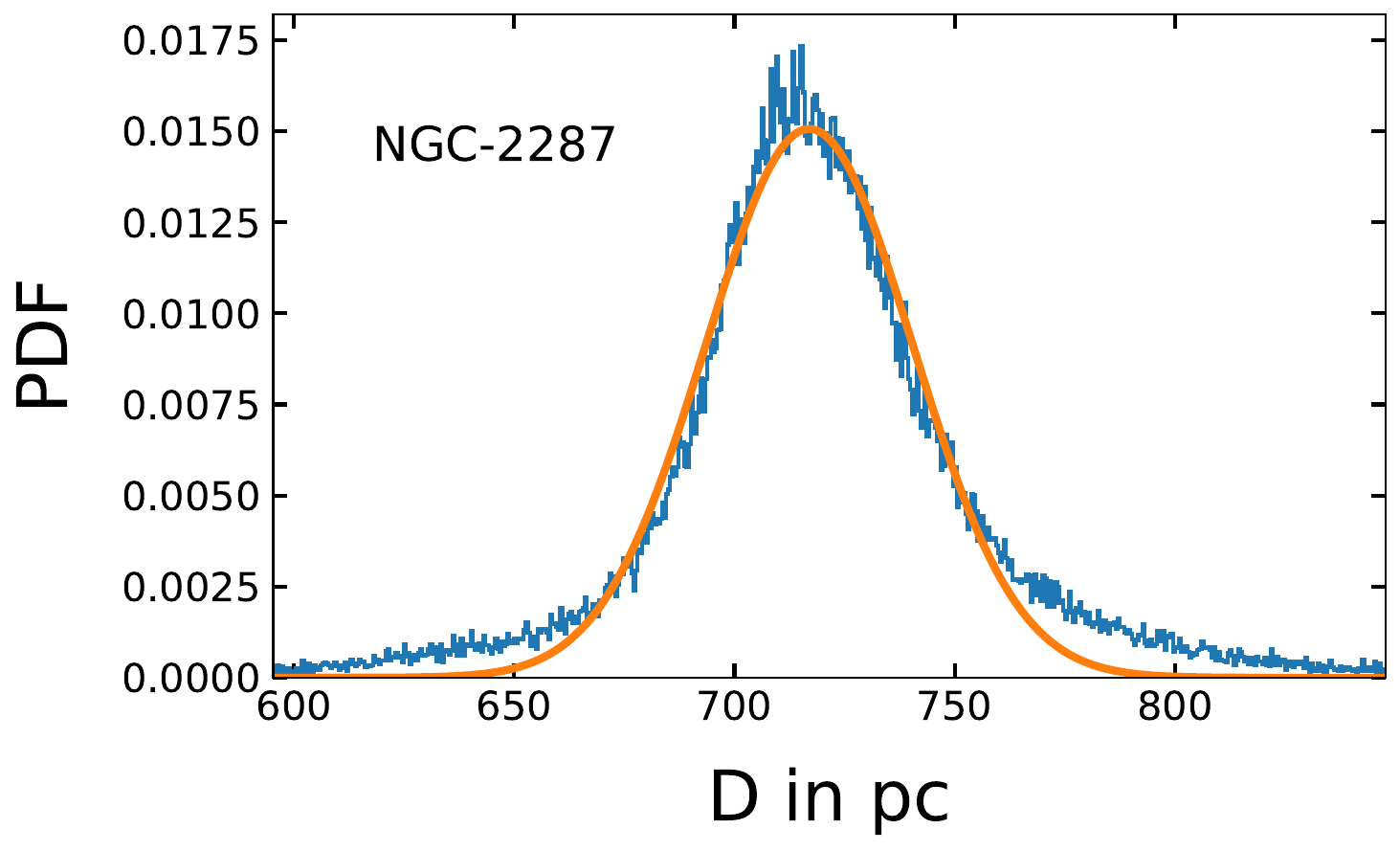}
\figsetgrpnote{Distance distribution of the cluster members for NGC 2287.}
\figsetgrpend

\figsetend

\begin{figure} 
\gridline{\fig{distance_fit_NGC-2287.pdf}{0.45\textwidth}{}}
\caption{
We estimate each cluster member's distance by cross-matching the data with \citet{ballierjones_2021}. The blue histogram shows the distance distribution of the cluster members we identify for NGC 2287 taking into account the estimated errors. The orange line shows the best-fit Gaussian to the distance distribution which we use as a prior to our estimation for the cluster's distance. }
\label{fig:cluster_distance}
\end{figure}
\subsection{Parameter estimation}
\label{S:cluster_props_params}
We use the publicly available Markov-Chain Monte Carlo (MCMC) toolkit \emcee\ \citep{emcee} to perform parameter estimation for each OC we analyse. In all cases, we use 128 walkers initially distributed according to the prior distributions. We perform 60,000 steps using the Metropolis-Hastings algorithm. We discard the first 30,000 steps as burn-in and trim the later steps by choosing 1 position in every 15 steps. We have tested for convergence in the posterior distributions and find that they converge well before 40,000 steps adopting the same burn-in and trimming (see \autoref{S:App-convergence_bayesian} and \autoref{fig:2287_corner_covergence} for more details). We have also tested varying $\pcluster$ and find that the posteriors remain statistically identical (see \autoref{S:App-pmemc} and \autoref{fig:comparison_mem_iso} for more details).   
\subsubsection{Source sampling}
\label{S:source_sampling}
In most of our clusters, the giant branch is not populated well enough. In general, $\age$ estimation from isochrone fitting using the giant branch can be tricky due to the rapid evolution of stars in the giant phase. Hence, we use up to the MS turn-off region to estimate the cluster properties. Within the MS, the brighter the source, the lower the photometric errors \citep{lindegren_2018}. On the other hand, as the magnitude decreases, the number of sources decrease fast because of the stellar mass function. As a result, if the actual observed sources, for example, within the single-star MS strip (\autoref{S:cluster_props_MS}), are sampled directly, the faint part of the CMD, populated by a large number of sources with high photometric errors, would receive a lot more weight in the calculation of likelihood and the MS turn-off region would receive very little weight. As a result, the likelihood calculation would become less sensitive to the MS turn-off region which usually has the most constraining power in determining the $\age$ of the cluster.     

We overcome these challenges by uniformly sampling 1000 \gaia-G values between magnitudes corresponding to the MS turn-off and G=18 for each OC. For each of these G values, we uniformly sample BP-RP within the range obtained from the spread we estimate for the single-star MS. These $\{$G, BP-RP$\}$ points represent our data for the single-star MS which we will denote as $\msdata$.
\subsubsection{Construction of likelihood}
\label{S:likelihood}
For each cluster, the parameters we want to estimate are $\age$, $\metal$, $D$, and $A_v$, which we collectively denote as $\vec{\theta}\equiv\{\age,\ \metal,\ D,\ A_v\}$. We generate isochrone models using the \mist\ isochrone package \citep{mist0,mist1}. While each isochrone in \mist\ takes $\age$, $\metal$, and $A_v$ as input, the distance $D$ is used to convert the absolute magnitudes generated by \mist\ \citep{mist0,mist1} to apparent magnitudes available in \gaia's DR3.   

We define the likelihood function ($\mathcal{L}$) as-
\begin{eqnarray}
    \ln\mathcal{L} & = & P(\vec{d} | M; \vec{\theta}) \nonumber\\
                   & = & -\frac{1}{2} \sum_i \left[ -\frac{(d_i - M_i(\vec{\theta}))^2}{s_i^2} + \ln (2\pi s_i^2) \right]
\end{eqnarray}
where, the subscript $i$ denotes the index of the data points, $M_i$ denotes the point on CMD given by \mist's model isochrones with parameters $\vec{\theta}$.
\begin{equation}
    s_i^2 = \sigma_i^2 + f^2,
\end{equation}
where, $\sigma_i$ is the error at $d_i$ estimated in \autoref{S:cluster_props_MS}.
and $f$ is an ad hoc numerical parameter, not associated with anything physical, introduced to stabilize the likelihood function \citep[for a detailed discussion, see][]{emcee}. 

Using the above likelihood function we proceed to evaluate the posterior distribution of $\vec{\theta}$. The posterior is defined as-
\begin{eqnarray}
    P(\vec{\theta} | \vec{d}, \vec{\sigma}; M) = P(\vec{d} | \vec{\theta}, \vec{\sigma}; M) P(\vec{\theta}).
\end{eqnarray}
where $P(\vec{\theta})$ denotes the prior distribution. 

\noprint{\figsetstart}
\noprint{\figsetnum{9}}
\noprint{\figsettitle{Corner plots showing two-dimensional joint and one-dimensional marginalized posterior distributions for cluster properties.}}

\figsetgrpstart
\figsetgrpnum{9.11}
\figsetgrptitle{NGC 2287}
\figsetplot{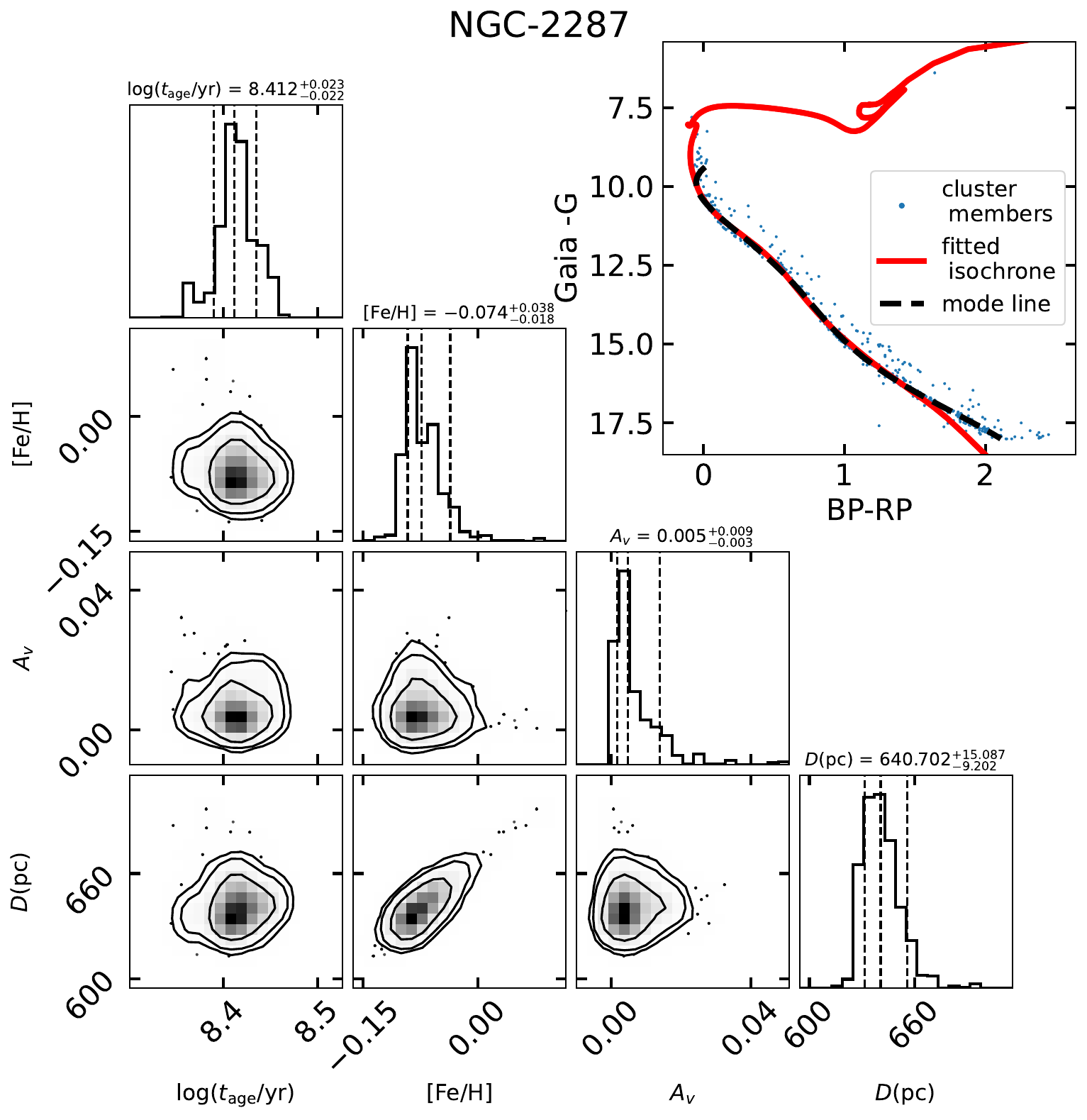}
\figsetgrpnote{Corner plot showing two-dimensional joint and one-dimensional marginalized posterior distributions for NGC 2287 cluster properties.}
\figsetgrpend

\figsetend

\begin{figure*}
\gridline{\fig{NGC-2287_24_04_24_corner_plots.pdf}{1.0\textwidth}{}}
\caption{Corner plot showing two-dimensional joint and one-dimensional marginalized posterior distributions for NGC-2287 cluster properties. The contours in the two-dimensional joint posterior distributions enclose $68\%$, $90\%$, and $95\%$ probabilities. The vertical lines in the one-dimensional posterior distributions show the median, and $16\%$ and $84 \%$ confidence bounds. We show the CMD in the right top corner. Dots represent cluster members, i.e., sources with $\pmem>\pcluster$. The black dashed and the red solid lines denote the best-fit line connecting the modes (see details in \autoref{S:cluster_props_MS}) and the best-fit isochrone created using the median values from the posterior distributions, respectively. Clearly, the isochrone is a very good fit for the CMD except near very low-G sources where \gaia ’s photometry has large errors and the theoretical isochrones have known issues. 
}
\label{fig:2287_corner_main_plot}
\end{figure*}
\subsubsection{Construction of priors}
\label{S:cluster_props_distance}
We use flat priors in $\age$, $\metal$, and $A_v$ in the range $10^{7}$--$10^{10}\,\yr$, $-1.0$--$0.5$, and $0$--$5$ magnitudes, respectively. 

The most straightforward way to estimate $D$ could have been inversion of $\plx$ given in \gaia's DR3. However, estimation of $D$, especially for stars fainter than $G=17$, via direct inversion of $\plx$ can introduce large errors due to uncertainties in \gaia's $\plx$ measurement. Moreover, direct inversion of negative $\plx$ is just incorrect. Instead, \citet{ballierjones_2021} implemented a Bayesian approach taking into account interstellar extinction, color, and magnitude to determine the posteriors for what they called the {\em photogeometric} distance from $\plx$. We cross-match the individual cluster members to the \citet{ballierjones_2021} catalog using RA-dec and cross-match radius $<1''$. We find that this choice of cross-match radius does not create multiple matches due to crowding in the clusters we have studied. Once we identify the sources in the catalog, we collect the median, 16th and 84th percentiles of the photogeometric distance posteriors given by \citet{ballierjones_2021}. For each cluster member, we approximate the photogeometric distance posterior as a Gaussian based on the median, 16th and 84th percentiles. We combine these to find the overall PDF. We then find the best-fit Gaussian for this combined PDF. This best-fit Gaussian is used as the prior distriution for $D$. 

\autoref{fig:2287_corner_main_plot} shows the posterior distributions for $D$, $\metal$, $A_v$, and $\age$ for our example OC NGC-2287. The three contours in the two-dimensional joint posterior distributions enclose $68\%$, $90\%$, and $95\%$ of the marginalised probability. The vertical dashed lines in the one-dimensional marginalised distributions from left to right show the $16$,  $50$, and $84$ percentiles. Top right panel shows the CMD for the cluster members (blue dots). We show the reference single-star MS identified using modes for the distribution of sources on the CMD (black dashed; \autoref{S:cluster_props_MS}) and the model isochrone created using \mist\ using the median of the posterior distribution of $\vec{\theta}$ (red solid). Clearly, our estimated parameters provide a good fit to the CMD. Interestingly, although the giant region was not considered in $\vec{d}$, the model isochrone fits the few giants present in NGC-2287 too. Also note that the model isochrone deviates from the reference single-star MS for G $\gtrsim17$. This is partially due to the relatively higher photometric errors for fainter sources. The under-prediction of magnitudes for stars in the mass range $0.25$ to $0.85\,\msun$ by \mist\ isochrones may be also be related to its limitations in the convection prescription \citep{brander2023}.

\tabcolsep3.2pt 
\begin{deluxetable*}{c|ccc|cccc|c|ccc|c}[h!]
\tablecaption{Global properties of all the clusters.}
\tabletypesize{\tiny}
\tablehead{
\colhead{Cluster} &\colhead{$\ncl$}&\colhead{$\nclcg$}&\colhead{$\nclhunt$}&\colhead{log($\age$/yr)}  &\colhead{$\metal$}& \colhead{Extinction} &\colhead{$D$} & \colhead{$D$} &\colhead{$\pmra$} &\colhead{$\pmdec$} &\colhead{$\plx$}&\colhead{$\rdstar$}  \\
\colhead{Name} &\colhead{($\pcluster$)}&\colhead{}&\colhead{}& 
\multicolumn{4}{c}{Posteriors from Bayesian parameter estimation} & \colhead{prior} &\colhead{} &\colhead{}&\colhead{} &\colhead{}\\
\colhead{} &\colhead{}&\colhead{}&\colhead{} &\colhead{}&\colhead{(dex)}& \colhead{(mag)} &\colhead{(pc)} & \colhead{(pc)}&\colhead{(m.a.s/yr)}&\colhead{(m.a.s/yr)}&\colhead{(m.a.s)}&\colhead{(pc)}
}
\startdata
Collinder-261 & $1910(0.93)$ & 1799 & 1335 & $9.666^{+0.036}_{-0.027}$ & $-0.486^{+0.115}_{-0.089}$ & $1.438^{+0.057}_{-0.068}$ & $2634^{+54}_{-66}$ & $2636 \pm 471$ & $-6.368^{+0.121}_{-0.119}$ & $-2.679^{+0.118}_{-0.119}$ & $0.345^{+0.06}_{-0.061}$ & 3 \\
Collinder-69$^\ddagger$ & $787(0.9)$ & 620 & 661 & $6.502^{+0.002}_{-0.001}$ & $0.019^{+0.008}_{-0.007}$ & $0.248^{+0.048}_{-0.043}$ & $396^{+3}_{-3}$ & $392 \pm 15$ & $1.172^{+0.517}_{-0.534}$ & $-2.111^{+0.319}_{-0.278}$ & $2.502^{+0.095}_{-0.094}$ & 1 \\
IC-4651 & $1360(0.97)$ & 806 & 693 & $9.239^{+0.006}_{-0.006}$ & $0.315^{+0.04}_{-0.036}$ & $0.19^{+0.028}_{-0.032}$ & $908^{+5}_{-6}$ & $912 \pm 44$ & $-2.434^{+0.23}_{-0.25}$ & $-5.037^{+0.243}_{-0.221}$ & $1.058^{+0.065}_{-0.094}$ & $<$1 \\
Melotte-101 & $810(0.95)$ & 472$^\dagger$ & 416 & $7.436^{+0.01}_{-0.012}$ & $-0.252^{+0.022}_{-0.029}$ & $1.326^{+0.005}_{-0.007}$ & $1728^{+21}_{-23}$ & $2027 \pm 186$ & $-6.33^{+0.108}_{-0.136}$ & $3.487^{+0.118}_{-0.114}$ & $0.456^{+0.063}_{-0.062}$ & 1 \\
Melotte-20$^\ddagger$ & $583(0.94)$ & 747 & 538 & $7.452^{+0.021}_{-0.021}$ & $0.041^{+0.018}_{-0.018}$ & $0.589^{+0.027}_{-0.027}$ & $180^{+4}_{-4}$ & $172 \pm 5$ & $23.073^{+0.821}_{-0.802}$ & $-25.349^{+0.99}_{-1.02}$ & $5.754^{+0.163}_{-0.181}$ & 4 \\
Melotte-22 & $1269(0.95)$ & 952 & 930 & $7.449^{+0.046}_{-0.03}$ & $-0.076^{+0.083}_{-0.035}$ & $0.659^{+0.036}_{-0.041}$ & $150^{+1}_{-3}$ & $135 \pm 3$ & $19.948^{+1.147}_{-1.177}$ & $-45.387^{+1.365}_{-1.341}$ & $7.361^{+0.192}_{-0.195}$ & 5 \\
NGC-1039 & $711(0.94)$ & 555 & 489 & $7.689^{+0.041}_{-0.005}$ & $0.14^{+0.017}_{-0.02}$ & $0.278^{+0.018}_{-0.009}$ & $533^{+8}_{-8}$ & $488 \pm 17$ & $0.66^{+0.268}_{-0.254}$ & $-5.806^{+0.283}_{-0.296}$ & $2.005^{+0.076}_{-0.088}$ & 2 \\
NGC-1647 & $688(0.94)$ & 604 & 504 & $7.839^{+0.031}_{-0.013}$ & $0.112^{+0.03}_{-0.03}$ & $1.308^{+0.009}_{-0.007}$ & $520^{+9}_{-9}$ & $575 \pm 20$ & $-1.084^{+0.253}_{-0.251}$ & $-1.539^{+0.252}_{-0.234}$ & $1.694^{+0.065}_{-0.068}$ & 2 \\
NGC-188 & $905(0.95)$ & 857 & 648 & $9.867^{+0.012}_{-0.011}$ & $0.2^{+0.035}_{-0.051}$ & $0.146^{+0.037}_{-0.03}$ & $1746^{+12}_{-16}$ & $1793 \pm 135$ & $-2.324^{+0.105}_{-0.105}$ & $-1.02^{+0.126}_{-0.121}$ & $0.525^{+0.038}_{-0.037}$ & 7 \\
NGC-2112 & $867(0.95)$ & 687 & 575 & $9.522^{+0.006}_{-0.006}$ & $-0.441^{+0.022}_{-0.026}$ & $2.258^{+0.013}_{-0.014}$ & $778^{+10}_{-8}$ & $1043 \pm 77$ & $-2.711^{+0.157}_{-0.139}$ & $4.282^{+0.138}_{-0.149}$ & $0.91^{+0.071}_{-0.054}$ & 3 \\
NGC-2287 & $679(0.96)$ & 625 & 508 & $8.412^{+0.023}_{-0.022}$ & $-0.074^{+0.038}_{-0.018}$ & $0.005^{+0.009}_{-0.003}$ & $641^{+15}_{-9}$ & $717 \pm 23$ & $-4.365^{+0.17}_{-0.168}$ & $-1.359^{+0.173}_{-0.173}$ & $1.359^{+0.046}_{-0.045}$ & 3 \\
NGC-2477 & $2696(0.94)$ & 1713 & 1773 & $8.95^{+0.007}_{-0.008}$ & $0.072^{+0.01}_{-0.008}$ & $1.024^{+0.015}_{-0.013}$ & $1263^{+15}_{-14}$ & $1384 \pm 66$ & $-2.424^{+0.159}_{-0.154}$ & $0.9^{+0.173}_{-0.163}$ & $0.689^{+0.041}_{-0.034}$ & 4 \\
NGC-2516 & $1874(0.95)$ & 652 & 1130 & $8.325^{+0.034}_{-0.05}$ & $0.227^{+0.08}_{-0.05}$ & $0.221^{+0.021}_{-0.021}$ & $414^{+19}_{-12}$ & $407 \pm 9$ & $-4.639^{+0.567}_{-0.523}$ & $11.211^{+0.445}_{-0.424}$ & $2.428^{+0.059}_{-0.064}$ & $<$1 \\
NGC-2539 & $570(0.96)$ & 485$^\dagger$ & 443 & $8.857^{+0.011}_{-0.01}$ & $-0.059^{+0.02}_{-0.027}$ & $0.191^{+0.012}_{-0.012}$ & $1176^{+17}_{-18}$ & $1253 \pm 69$ & $-2.326^{+0.103}_{-0.115}$ & $-0.551^{+0.104}_{-0.091}$ & $0.757^{+0.049}_{-0.046}$ & 2 \\
NGC-2632 & $730(0.93)$ & 685 & 665 & $8.808^{+0.02}_{-0.014}$ & $0.267^{+0.007}_{-0.006}$ & $0.004^{+0.005}_{-0.003}$ & $174^{+1}_{-1}$ & $183 \pm 4$ & $-35.911^{+1.06}_{-1.067}$ & $-12.884^{+0.915}_{-0.961}$ & $5.409^{+0.115}_{-0.115}$ & 5 \\
NGC-2682 & $1178(0.93)$ & 598 & 743 & $9.752^{+0.004}_{-0.004}$ & $0.088^{+0.006}_{-0.008}$ & $0.003^{+0.004}_{-0.002}$ & $789^{+3}_{-4}$ & $835 \pm 34$ & $-10.973^{+0.206}_{-0.175}$ & $-2.919^{+0.208}_{-0.181}$ & $1.152^{+0.053}_{-0.053}$ & 6 \\
NGC-6124 & $1501(0.95)$ & 1273 & 930 & $8.481^{+0.009}_{-0.011}$ & $0.134^{+0.017}_{-0.013}$ & $2.11^{+0.008}_{-0.007}$ & $501^{+6}_{-4}$ & $613 \pm 23$ & $-0.224^{+0.318}_{-0.329}$ & $-2.106^{+0.322}_{-0.334}$ & $1.594^{+0.07}_{-0.074}$ & 1 \\
NGC-6819 & $1689(0.95)$ & 1527 & 937 & $9.694^{+0.029}_{-0.016}$ & $-0.094^{+0.012}_{-0.004}$ & $0.406^{+0.018}_{-0.028}$ & $2056^{+24}_{-30}$ & $2465 \pm 268$ & $-2.893^{+0.098}_{-0.097}$ & $-3.864^{+0.108}_{-0.107}$ & $0.375^{+0.043}_{-0.039}$ & 4 \\
NGC-6939 & $821(0.96)$ & 636 & 559 & $9.471^{+0.005}_{-0.007}$ & $-0.842^{+0.014}_{-0.014}$ & $1.567^{+0.009}_{-0.01}$ & $1422^{+1}_{-1}$ & $1804 \pm 124$ & $-1.819^{+0.103}_{-0.113}$ & $-5.463^{+0.104}_{-0.097}$ & $0.525^{+0.035}_{-0.035}$ & 3 \\
NGC-6940 & $897(0.95)$ & 572 & 453 & $8.894^{+0.017}_{-0.022}$ & $0.025^{+0.041}_{-0.041}$ & $0.83^{+0.023}_{-0.022}$ & $1061^{+36}_{-25}$ & $1011 \pm 43$ & $-1.944^{+0.199}_{-0.218}$ & $-9.431^{+0.238}_{-0.233}$ & $0.95^{+0.066}_{-0.336}$ & $<$1 \\
NGC-7789 & $2830(0.92)$ & 2897 & 2158 & $9.605^{+0.012}_{-0.007}$ & $-0.742^{+0.028}_{-0.016}$ & $1.001^{+0.015}_{-0.017}$ & $1421^{+13}_{-16}$ & $1949 \pm 164$ & $-0.912^{+0.116}_{-0.116}$ & $-1.949^{+0.131}_{-0.111}$ & $0.478^{+0.041}_{-0.042}$ & 4 \\
\enddata
\tablecomments{
\scriptsize
Cluster properties including $\age$, $\metal$, $D$, $\reddening$, $\pmra$, $\pmdec$, $\plx$, $\rdstar$, and $\ncl$ for the 21 OCs we have studied. $\rdstar$ is the $\rproj$ where the surface density of cluster members $\sigmacluster=\sigmafield$, the average surface density of all sources with $15\leq\rproj/\pc\leq20$. For comparison, we also show $\nclcg$ ($\nclhunt$), the number of cluster members identified by \cgmem\ with probability $>0.7$ (the members identified by \hunmem\, which are within the tidal radius of the cluster and $G$ $<$ 18 mag). The central value and the errors for each property denote the median, and $16\%$ and $84\%$ credible intervals. In most OCs we find significantly higher member numbers compared to those identified in \cgmem\ and \hunmem\ (see discussion in \autoref{s:results_ncl}). Although, \cgmem\ estimated fewer than $500$ members for Melotte-101 and NGC-2539, we have analyzed these OCs and find $\ncl>500$ (marked by $^\dagger$). Clusters where properties show significant multi-modal distributions are marked with `$\ddagger$' for the corresponding properties.
Collinder-69 and Melotte-20 show multi-modal distributions for $\pmra$. The dominant $\pmra$/(m.a.s/yr) modes are 0.95 and 1.53 for Collinder-69 and 22.65 and 23.28 for Melotte-20.
}
\label{tab:cluster_properties_table} 
\end{deluxetable*}
\section{Results and Discussion}
\label{S:results}
We have summarised all estimated properties for the 21 OCs in   \autoref{tab:cluster_properties_table}. For each OC, we have listed the number of cluster members according to our analysis. We also report the median and the spread between 
the $16$th and $84$th percentiles for the astrometric and cluster properties for each OC. Overall, the OCs in our sample exhibit a reasonably large range in properties. The number of 
cluster members $\ncl$ varies from 570 for NGC-2539 to 2830 for NGC-7789 with a median $\ncl=884$. These clusters are estimated to be as young as $\age=10\,\Myr$ (for Collinder 69) to as old as $\age=7\,\Gyr$ (for NGC-188) with a median at $\age=724\,\Myr$ (for NGC-2539). The most distant and the closest OCs among the 21 are Collinder-261 at $D/\pc\approx2.7\times10^3$ and Mellotte 22 at $D/\pc\approx1.5\times10^2$. Our clusters exhibit a range $-0.84\leq\metal\leq0.31$ spanning both sides of the solar metallicity. We estimate that $A_v$ ranges from $0.003$ to $2.26$ with a median at $0.59$.     

\noprint{\figsetstart}
\noprint{\figsetnum{10}}
\noprint{\figsettitle{Cumulative member counts as a function of projected distance.}}

\figsetgrpstart
\figsetgrpnum{10.16}
\figsetgrptitle{NGC 2682}
\figsetplot{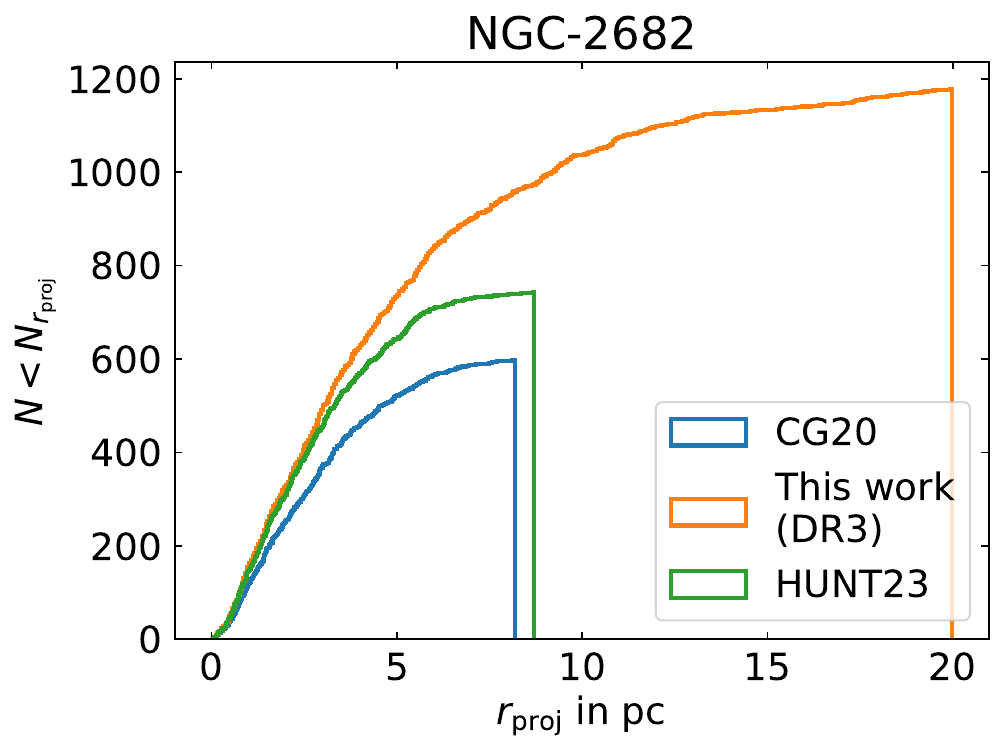}
\figsetgrpnote{Cumulative member count as a function of projected distance for NGC 2682.}
\figsetgrpend

\figsetgrpstart
\figsetgrpnum{10.17}
\figsetgrptitle{NGC 6124}
\figsetplot{radial_density_NGC-6124_v1.pdf}
\figsetgrpnote{Cumulative member count as a function of projected distance for NGC 6124.}
\figsetgrpend

\figsetgrpstart
\figsetgrpnum{10.18}
\figsetgrptitle{NGC 6819}
\figsetplot{radial_density_NGC-6819_v1.pdf}
\figsetgrpnote{Cumulative member count as a function of projected distance for NGC 6819.}
\figsetgrpend

\figsetgrpstart
\figsetgrpnum{10.19}
\figsetgrptitle{NGC 6939}
\figsetplot{radial_density_NGC-6939_v1.pdf}
\figsetgrpnote{Cumulative member count as a function of projected distance for NGC 6939.}
\figsetgrpend

\figsetgrpstart
\figsetgrpnum{10.20}
\figsetgrptitle{NGC 6940}
\figsetplot{radial_density_NGC-6940_v1.pdf}
\figsetgrpnote{Cumulative member count as a function of projected distance for NGC 6940.}
\figsetgrpend

\figsetgrpstart
\figsetgrpnum{10.21}
\figsetgrptitle{NGC 7789}
\figsetplot{radial_density_NGC-7789_v1.pdf}
\figsetgrpnote{Cumulative member count as a function of projected distance for NGC 7789.}
\figsetgrpend

\figsetend

\begin{figure}
\gridline{\fig{radial_density_NGC-2682_v1.pdf}{0.45\textwidth}{}}
\caption{Cumulative member count as a function of $\rproj$ for NGC 2682 as an example. Orange and blue denote the cumulative member counts from our analysis and \hunmem\ using \gaia's DR3 and those from \cgmem\ using \gaia's DR2. We not only identify more members at higher distances from the fiducial cluster centre, but we also identify more members compared to those identified by \cgmem\ and \hunmem\ even within $\rproj/\pc<8$ where they analyzed.
}
\label{fig:special_2682_cdf}
\end{figure}

\noprint{\figsetstart}
\noprint{\figsetnum{11}}
\noprint{\figsettitle{Histograms of relative errors for identified cluster members.}}

\figsetgrpstart
\figsetgrpnum{11.16}
\figsetgrptitle{NGC-2682}
\figsetplot{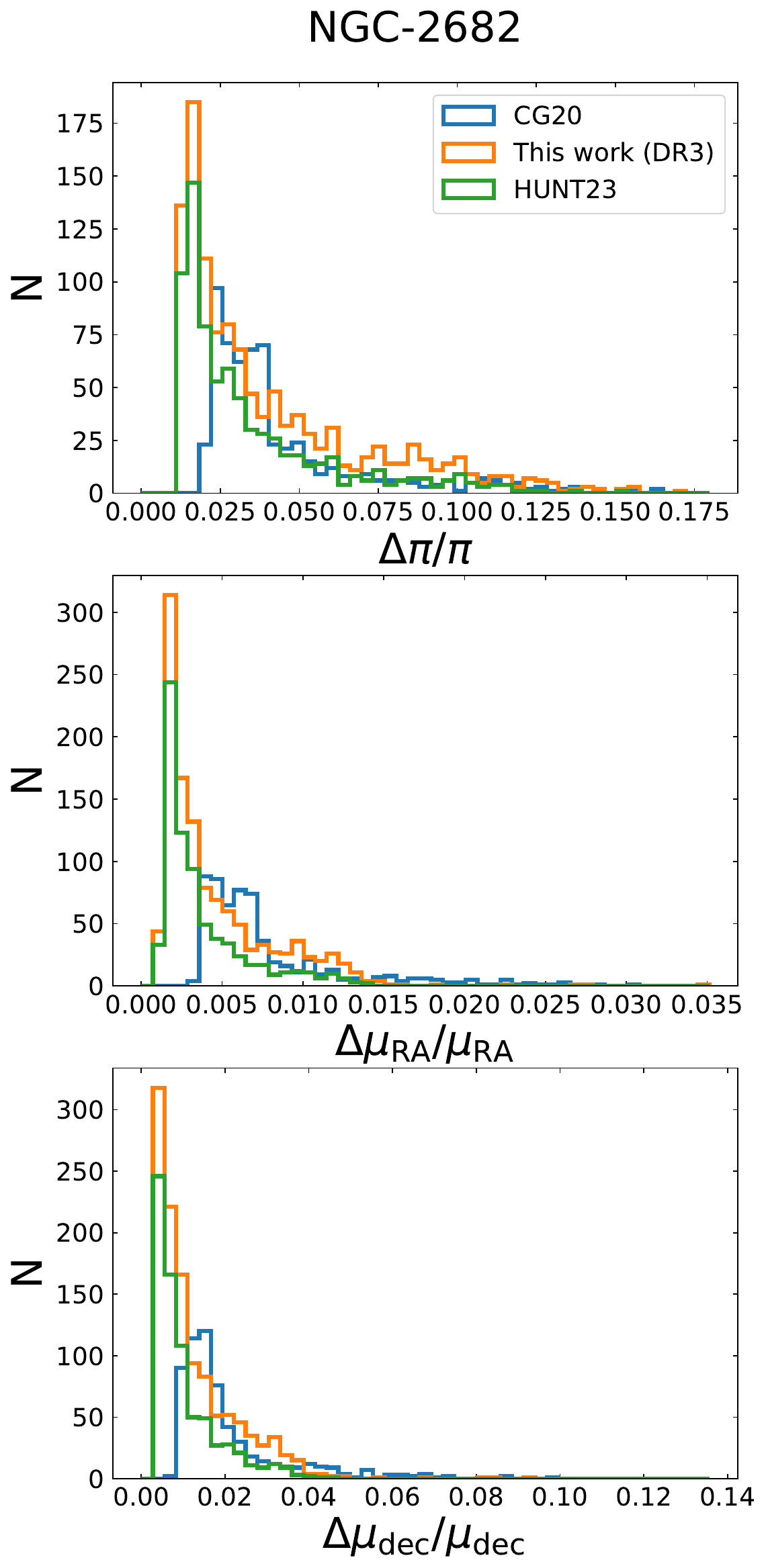}
\figsetgrpnote{Histograms of relative errors for identified cluster members of NGC-2682.}
\figsetgrpend

\figsetgrpstart
\figsetgrpnum{11.17}
\figsetgrptitle{NGC-6124}
\figsetplot{NGC-6124_astrometric_com_v2.pdf}
\figsetgrpnote{Histograms of relative errors for identified cluster members of NGC-6124.}
\figsetgrpend

\figsetgrpstart
\figsetgrpnum{11.18}
\figsetgrptitle{NGC-6819}
\figsetplot{NGC-6819_astrometric_com_v2.pdf}
\figsetgrpnote{Histograms of relative errors for identified cluster members of NGC-6819.}
\figsetgrpend

\figsetgrpstart
\figsetgrpnum{11.19}
\figsetgrptitle{NGC-6939}
\figsetplot{NGC-6939_astrometric_com_v2.pdf}
\figsetgrpnote{Histograms of relative errors for identified cluster members of NGC-6939.}
\figsetgrpend

\figsetgrpstart
\figsetgrpnum{11.20}
\figsetgrptitle{NGC-6940}
\figsetplot{NGC-6940_astrometric_com_v2.pdf}
\figsetgrpnote{Histograms of relative errors for identified cluster members of NGC-6940.}
\figsetgrpend

\figsetgrpstart
\figsetgrpnum{11.21}
\figsetgrptitle{NGC-7789}
\figsetplot{NGC-7789_astrometric_com_v2.pdf}
\figsetgrpnote{Histograms of relative errors for identified cluster members of NGC-7789.}
\figsetgrpend

\figsetend

\begin{figure}
\gridline{\fig{NGC-2682_astrometric_com_v2.pdf}{0.45\textwidth}{}}
\caption{Histograms of relative errors in $\plx$ (top), $\pmra$ (middle), and $\pmdec$ (bottom) for the cluster members we identify (orange) and those identified by \hunmem\ (green) using \gaia's DR3. We also show the corresponding distributions for \cgmem\ (blue) which used \gaia's DR2. 
The shift in the peak of the distributions towards lower errors in \hunmem\ and this analysis is due to the improvement of astrometric measurements in \gaia's DR3.}
\label{fig:astrometric_2_error_NGC_2682}
\end{figure}
\subsection{Number of cluster members}
\label{s:results_ncl}
The cluster members we find can be directly compared with those mentioned in \cgmem\ with some caveats. There are several key differences between these two studies. For example, we have used data from \gaia's DR3, while \cgmem\ used that from DR2. The meaning and method of estimating $\pmem$ are also very different between these studies, as a result, a meaningful comparison cannot adopt the same $\pcluster$ used in these studies. Instead, we compare our membership numbers with those reported in \cgmem\,(\autoref{tab:cluster_properties_table}). 
We find that in 19 out of  21 clusters we have studied, the cluster member $\ncl$ identified by us is higher than that in \cgmem\ ($\nclcg$). We find that the median of $\ncl/\nclcg$ using $\pmemcg>0.7$ ($0.5$) is $\approx1.3$ ($1.2$), whereas, the largest (smallest) $\ncl/\nclcg\approx2.9$ ($0.8$) for NGC-2516 (Mellotte-20) adopting $\pmemcg>0.7$.        

A detailed analysis sheds light on the potential reasons for these differences. One possible source of this difference could have been the difference in the considered limiting $\rproj$. As an example, we show the cumulative source count as a function of $\rproj$ in \autoref{fig:special_2682_cdf} for the rich, old, and extremely well-studied cluster NGC-2682 (M67). While \cgmem\ stops identifying cluster members at $\rproj/\pc\leq 8$, we continue finding members all the way to $\rproj/\pc=20$, although the rate of increase in source count significantly decreases beyond $\rproj/\pc\sim15$. Interestingly, even within $\rproj/\pc=8$, we identify more members compared to \cgmem\ by a factor of $\approx2$. Thus, our choice of $\rproj$ is not the primary source of the difference in $\ncl$. We find that this difference is also {\em not} due to the fact that \cgmem\ used DR2 and we, DR3. Even with the DR2 data, our analysis identifies almost the same cumulative source counts as a function of $\rproj$ in all clusters. Nevertheless, because of the much-improved uncertainties in DR3, the cluster members we identify naturally exhibit significantly lower fractional errors in $\{\plx, \promo\}$ compared to those identified by \cgmem\ (\autoref{fig:astrometric_2_error_NGC_2682}). We have tested whether the astrometric parameters, distance, magnitude, and color for the cluster members in our study show any systematic differences compared to these properties for the members identified by \cgmem\ (\autoref{fig:special_2682_comparison}). We find that the $D$, $\promo$ show very similar distributions in both studies, our study simply finds a higher number of members. This implies that the new members we find are not astrometric outliers or coming from the outskirts of the cluster and bolsters our belief that they are genuine members. Nevertheless, a close inspection of the CMDs reveals that \cgmem\ systematically missed a higher fraction of relatively fainter ($G\gtrsim15$) binary MS sources. MS binaries are of course interesting since they take part in formation of stellar exotica through binary stellar evolution and stellar dynamics  \citep{Moe_2017,Wang_2020,Mohandasan_2024,el_badry_binary_review_2024}. It also appears that \cgmem\ missed many sources in the blue straggler region. Blue stragglers are of course very interesting dynamical tracers in a star cluster \citep{knigge2015,ferraro_2023} and if they are there, it is desirable for a member-finding algorithm to identify them. Our member-finding strategy does not depend on the proximity of the sources to any well-defined sequences on the CMD, including the MS. As a result, this strategy may be better suited to identify stragglers. 

Similar to the above example, in all of the 21 OCs, the astrometric errors in the members we find are significantly lower because of the shift from DR2 to DR3. Moreover, in all cases, we find that the astrometric properties and $D$ found by this study and \cgmem\ show very similar distributions, peaking roughly at the same values. In 19 of the 21 OCs we have analysed, the comparison of $\ncl$ shows exactly the same trends, i.e., we identify more members and {\em all} members of \cgmem\ are also identified as members by us. The case for Mellotte 20 is somewhat different. In this case, the limiting $\rproj$ considered by \cgmem\ is $30\,\pc$, significantly higher than ours. As a result, $\ncl/\nclcg<1$ for Melotte-20. 

In another substantial recent study \hunmem\ have used a density-based algorithm HDBSCAN to identify cluster members, which they later verified using a CMD classifier based on BNN. They used the DR3 data and as a result, it is very interesting to compare our results with theirs as well. \hunmem\ used stars up to $G=20$ compared to a cut-off of $G=18$ used in our analysis. For an apples-to-apples comparison, we compare their `good' members\footnote{Members within the tidal radius of the cluster as estimated in \hunmem.} up to $G=18$, with those identified by ours (\autoref{fig:special_2682_cdf}--\ref{fig:special_2682_comparison}).  
As expected, the astrometric errors used by \hunmem\ match well with ours, so does the peaks in the distributions for the cluster properties. We find that \hunmem\ also stopped identifying cluster members at $\rproj/\pc\leq8$ similar to \cgmem\ . However, \hunmem\ found more members compared to \cgmem\ and fewer members compared to our study (by a factor of 1.5) within that region for the example cluster NGC-2682. Using the above-mentioned adjustments for a fair comparison, we find that the ratio between the number of members found by \hunmem\ ($\nclhunt$) varies from $\ncl/\nclhunt=1.08$ for cluster Melotte-20 to $1.98$ for cluster NGC-6940, with a median $1.45$ for cluster NGC-1039. \autoref{fig:all_violin_plot} shows a comparison of $\ncl$ estimated in this study with those given in \cgmem\ and \hunmem\ for all clusters. Interestingly, in the case of NGC-2516, $\ncl$ estimated by us matches well with the estimate of \hunmem\, both of which are almost three times higher than the estimate given in \cgmem. Although to a much lesser extent compared to \cgmem, we find that \hunmem\ also have a higher chance of missing low-mass binary and blue-straggler members.   
\startlongtable
\tabcolsep9.0pt 
\begin{deluxetable*}{c|cccccccc}
\tablecaption{List of individual cluster members. }
\scriptsize
\tablehead{\colhead{Cluster Name}&\colhead{\ra} &
\colhead{\dec}&\colhead{$\pmra$}&\colhead{$\pmdec$}&\colhead{$\plx$} &\colhead{G}& \colhead{BP-RP} &\colhead{$\pmem$}  \\
\colhead{} &\colhead{(deg)} &\colhead{(deg)} &\colhead{(mas\ $\yr^{-1})$}&\colhead{(mas\ $\yr^{-1})$)}&\colhead{(mas)} &\colhead{(mag)}& \colhead{(mag)} &\colhead{}   
}
\startdata
NGC-&102.42 & -22.08 & -4.48 &  -1.30 &      1.37 &            14.74 &   1.00 &                1.00  \\
2287&102.57 & -22.04 & -4.69 &  -0.67 &      1.88 &            16.68 &   1.86 &                0.99  \\
$\pmem > \pcluster = 0.96$&101.28 & -22.12 & -4.57 &  -1.99 &      0.96 &            16.84 &   1.30 &                1.00\\ 
& \vdots&  \vdots& \vdots&   \vdots&     \vdots&          \vdots & \vdots &  \vdots \\ 
\hline
NGC-&101.10 & 22.32  &  -0.35 &      0.67 & 0.06&            16.40 &   1.39 &                0.00 \\
2287&101.10 & 22.32  &  -0.28 &      5.43 &  0.60 &         17.20 &   1.15 &                0.00  \\
$\pmem < \pcluster = 0.96$&101.10 & 22.32 & -1.68 &  6.95 &      0.47 &            15.84 &   0.79 &                0.00 \\
& \vdots&  \vdots& \vdots&   \vdots&     \vdots&          \vdots & \vdots &  \vdots \\ 
\hline
\hline
NGC-&133.69 & 11.02 & -11,12 &  -2.91 &     1.10 &            14.17 &   0.82 &                1.00  \\
2682&133.46 & 11.10 & -10.93 &  -2.31 &     1.20 &            12.87 &   0.76 &                1.00  \\
$\pmem > \pcluster = 0.93$&133.52 & 11.35 & -10.84 &  -2.65 &     1.10 &            15.45 &   1.14 &                1.00 \\
& \vdots&  \vdots& \vdots&   \vdots&     \vdots&          \vdots & \vdots &  \vdots \\ 
\hline
NGC-&133.64 & 10.81 & -33.56 &  2.57 &      3.15 &            12.66 &   0.89 &                0.00 \\
2682&133.64 & 10.81 & -33.11 &  3.08 &      3.48 &            9.28 &   0.17 &                0.00  \\
$\pmem < \pcluster = 0.93$&133.64 & 10.80 & 0.73 &  -8.14 &      0.88 &            17.07 &   1.30 &                0.00 \\
& \vdots&  \vdots& \vdots&   \vdots&     \vdots&          \vdots & \vdots &  \vdots \\ 
\hline
\hline
\vdots& \vdots&  \vdots& \vdots&   \vdots&     \vdots&          \vdots & \vdots &  \vdots \\ 
\enddata
\tablecomments{
\scriptsize
Properties including RA, dec, $\pmra$, $\pmdec$, $\plx$, \gaia-G, BP-RP, $\pmem$ of all analyzed sources within $\rproj/\pc<20$ for the 21 OCs we have studied 20 pc region of $21$ clusters. For each OC, we also mention the critical $\pmem$, $\pcluster$ above which sources are considered cluster members in this study. The full table has not yet been made public, but will be shared upon reasonable request to the authors.}
\label{tab:all_star_table}
\end{deluxetable*}
\autoref{tab:all_star_table} provides a truncated summary for the astrometric and photometric properties of all sources we have analysed within $\rproj/\pc\leq20$ from the 21 clusters. For all sources, independent of whether they are deemed members in our study or not, we provide position, magnitude, color, astrometric properties, our estimated $\pmem$, and the $\pcluster$ we have used for each cluster. Since we provide the estimated $\pmem$ for all sources independent of whether we consider them members or not, future studies can easily choose to use a different $\pcluster$ if so desired. The complete version of the table has not yet been made public, but will be shared upon reasonable request to the authors.

\noprint{\figsetstart}
\noprint{\figsetnum{12}}
\noprint{\figsettitle{Comparisons of properties of cluster members.}}

\figsetgrpstart
\figsetgrpnum{12.16}
\figsetgrptitle{NGC 2682}
\figsetplot{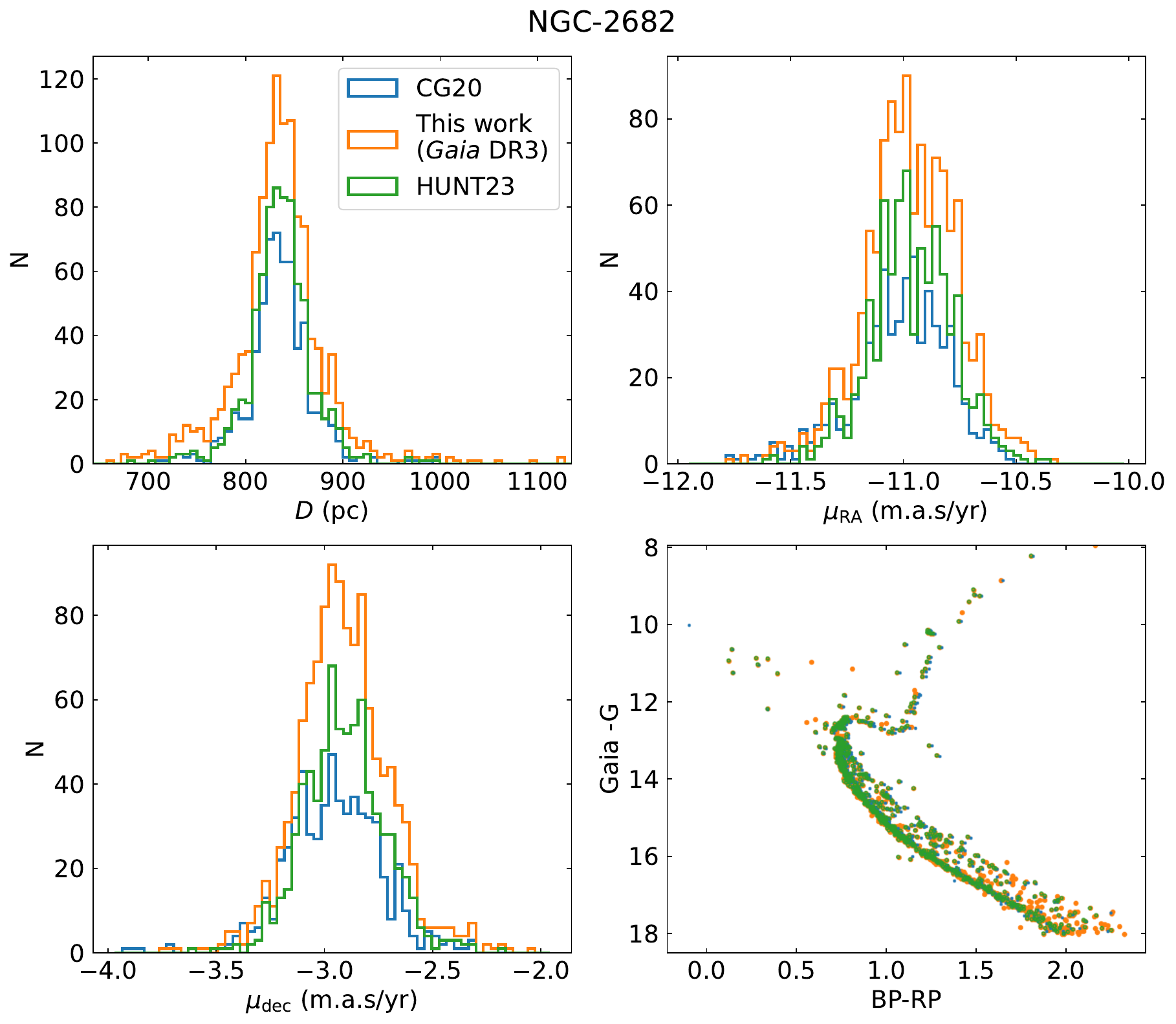}
\figsetgrpnote{Comparison of the properties of cluster members identified by this study, HUNT23, and CG20 for NGC 2682.}
\figsetgrpend

\figsetgrpstart
\figsetgrpnum{12.17}
\figsetgrptitle{NGC 6124}
\figsetplot{NGC-6124_com_hunt.pdf}
\figsetgrpnote{Comparison of the properties of cluster members identified by this study, HUNT23, and CG20 for NGC 6124.}
\figsetgrpend

\figsetgrpstart
\figsetgrpnum{12.18}
\figsetgrptitle{NGC 6819}
\figsetplot{NGC-6819_com_hunt.pdf}
\figsetgrpnote{Comparison of the properties of cluster members identified by this study, HUNT23, and CG20 for NGC 6819.}
\figsetgrpend

\figsetgrpstart
\figsetgrpnum{12.19}
\figsetgrptitle{NGC 6939}
\figsetplot{NGC-6939_com_hunt.pdf}
\figsetgrpnote{Comparison of the properties of cluster members identified by this study, HUNT23, and CG20 for NGC 6939.}
\figsetgrpend

\figsetgrpstart
\figsetgrpnum{12.20}
\figsetgrptitle{NGC 6940}
\figsetplot{NGC-6940_com_hunt.pdf}
\figsetgrpnote{Comparison of the properties of cluster members identified by this study, HUNT23, and CG20 for NGC 6940.}
\figsetgrpend

\figsetgrpstart
\figsetgrpnum{12.21}
\figsetgrptitle{NGC 7789}
\figsetplot{NGC-7789_com_hunt.pdf}
\figsetgrpnote{Comparison of the properties of cluster members identified by this study, HUNT23, and CG20 for NGC 7789.}
\figsetgrpend

\figsetend

\begin{figure}
\gridline{\fig{NGC-2682_com_hunt.pdf}{0.45\textwidth}{}}
\caption{Comparison of the properties of cluster members identified by this study (orange), \hunmem\ (green), and \cgmem\ (blue) for NGC 2682 as an example. Although we find significantly higher number of cluster members, the members are well-centered in $D$, $\pmra$, and $\pmdec$. Comparison of the CMDs denotes that we identify several new blue stragglers and many more low-mass stellar binaries compared to those found by \hunmem\ and \cgmem\ in NGC-2682.}
\label{fig:special_2682_comparison}
\end{figure}
\subsection{Cluster properties}
\label{s:results_properties}
Several past studies have provided estimates for the basic cluster properties using various methods. Keeping true to the spirit of this study, we intentionally choose studies that did not focus on specific clusters. \citet{dias_2002} released a catalogue of 1537 OCs and their properties, including coordinates, $\age$, apparent diameter, colour excess, $D$, astrometric properties, and $\metal$ based primarily on WEBDA \citep{lauberts1982} and ESO \citep{lynga1987} catalogues. It is one of the earliest efforts to collect a large number of clusters and their properties inside one single catalogue, which over time became a very useful reference to easily find the properties of a large number of OCs. However, due to the inclusion of data from different surveys, it is difficult to understand the systematics. Indeed, identifying databases obtained from homogeneous analysis of similar quality data can be challenging. For example, while \citet{netopil_2016} estimated $\metal$ for a large number of OCs using spectroscopy, their reported $\age$ was simply estimated by taking the mean of a variety of different past studies. In order to avoid such complications, we carefully investigate past studies and compare our results only with those found by recent studies employing {\em nearly} homogeneous analysis of similar quality data. Even in recent studies, care is needed to identify the truly independent self-consistent estimates of OC properties. For example, \citet{Bossini2019} did not use $\metal$ as a free parameter in their estimations. Instead, they used fixed metallicity values, $\metal=0$, or values taken directly from \citet{netopil_2016}. On the other hand, \cgmem\ results are based on machine learning where the artificial neural network was trained using OC properties reported in \citet{Bossini2019}. In a different study based on \gaia\, DR2 data, \citet{dias2021} homogeneously estimated cluster properties of  1743 clusters. They considered cluster members from previous studies and evaluated the cluster properties using isochrone fitting following the formalism similar to \citet{monterio2020}. Most recently, \hunmem\ estimated OC properties using a BNN, where training sets for the CMD classifier were developed based on theoretical isochrones. Hence, we restrict our comparisons to the results of \citet{dias2021} for all four cluster properties. We compare all properties except $\metal$ with those found in \citet{Bossini2019}, \cgmem, and \hunmem. We also compare $\metal$ with the spectroscopic estimates of \citet{netopil_2016} wherever available. Since the property estimates came from machine learning in several of these studies, errorbars are often absent. In case of \citet{dias2021}, the estimates represent the mean and $1\sigma$ errorbars estimated using bootstrap. 

\begin{figure*}[h!]
\gridline{\fig{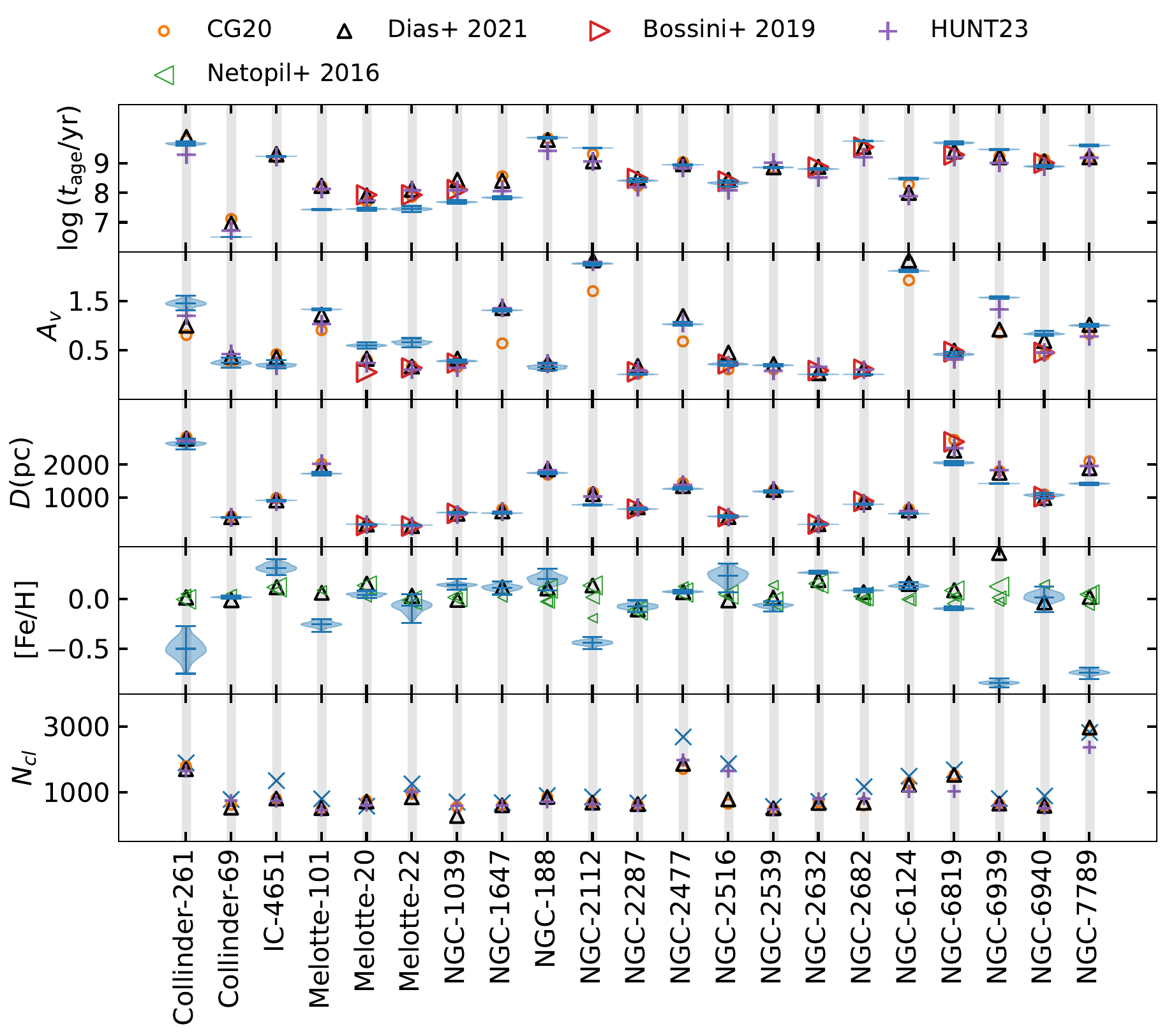}{1.01\textwidth}{}}
\caption{Comparison of cluster properties we estimate with those found in selected past studies \citep[\cgmem,][]{netopil_2016,Bossini2019,dias2021,hunt_2023}. Cluster names are shown along the horizontal axis. Orange dot, red triangle, violet `+', black triangle, and green triangle represent values from \cgmem, \citet{Bossini2019}, \hunmem\ , \citet{dias2021}, and \citet{netopil_2016} respectively. From small to large, the increasing size of the green triangles signifies $\metal$ estimated using photometry, low-resolution, and high-resolution spectroscopy. Blue violins and the horizontal lines denote the posterior distributions, and the median, 5th and 95th percentiles from our Bayesian parameter estimation.} 

\label{fig:all_violin_plot}
\end{figure*}

\autoref{fig:all_violin_plot} shows a detailed comparison of the cluster properties between our estimates and those in the past studies mentioned above. Where applicable, we show the detailed posterior distributions obtained in our studies (blue violins);
the horizontal lines in each violin denote the median, 5th, and 95th percentiles for the posterior distributions. Our estimated $\age$ match well with those estimated in previous studies. 
 
Our estimates for $\reddening$ match well with those of \citet{dias2021}, \hunmem, and \citet{Bossini2019}, where available. There are small differences between our estimates with those in \cgmem\ for some OCs.  
NGC-1647, NGC-2112, and Collinder 261 show the highest levels of mismatch with \cgmem. In all of these cases, our estimated $\reddening$ is higher compared to the estimate of \cgmem. We notice that these OCs are within $\pm<16^\circ$ to the galactic plane. Hence, a high level of $\reddening$ is not surprising. Interestingly, our estimates match reasonably well with those in \citet{dias2021} and \hunmem\ even for these OCs.

Our $D$ estimates show a reasonably good match with all previous studies. Nevertheless, for NGC 6819, NGC 6939, and NGC 7789 our estimates of $D$ are somewhat lower compared to the estimates of \cgmem\ and \hunmem\,. 

As mentioned earlier, \metal\ is one of the least addressed topics in past studies. In most cases, it is kept fixed at solar value or previously known values from \citet{netopil_2016}. The fourth row from the top of \autoref{fig:all_violin_plot}, shows the $\metal$ estimates by \citet{netopil_2016}. From the smallest to the largest, the green triangles denote photometric, low-resolution (LR), and high-resolution (HR) spectroscopic $\metal$ measurements. Our \metal\ estimates match well with the estimates of \citet{netopil_2016} for most OCs. Notable exceptions are some old OCs such as Collinder-261, NGC-6939, and NGC-7789, with $\log(\age/\yr)>9.5$. For these OCs, our \metal\ values are lower compared to those given in \citet{netopil_2016}. In general, older OCs are expected to have lower $\metal$ \citep[e.g.,][]{bergemann_2014}. 
Further investigation reveals that the $\metal$ estimates in \citet{netopil_2016} for OCs NGC-6939 (NGC-7789) using HR spectra were based on only 1 (5) star(s), thus, may not be robust. We encourage collection of more HR spectra for more reliable $\metal$ estimates for these old OCs. \citet{dias2021} used priors for \metal\ motivated from \citet{monterio2020}. \citet{monterio2020} used values and uncertainties directly from HR spectroscopic studies such as \citet{netopil_2016}, if available. Otherwise, they relied on the Galactic metallicity gradient values from \citet{Donor_2020}. In most of the cases, we find a good match between our estimates of $\metal$ and the estimates of \citet{dias2021}. 

\section{CONCLUSION}
\label{S:conclusion}
We have developed a non-parametric procedure to identify members of star clusters based on the astrometric parameters found by \gaia. In the membership analysis we self-consistently take into account the correlated errors in the astrometric parameters for each source. Once the members are identified, we utilize them to estimate global cluster properties including $\age$, $\metal$, $D$, and $\reddening$ via isochrone fitting. We have developed a method that eliminates the need for human intervention and subjectivity during isochrone fitting. As a first trial, we have applied these techniques for $21$ relatively rich ($\ncl>500$) and nearby ($D<3.0\,\kpc$) clusters identified by \cgmem\ (\autoref{S:data_selection}). We collect \gaia\ DR3 data from the previously known cluster centers up to $\rproj/\pc=20$ for the $21$ OCs and estimate the membership probabilities $\pmem$ for each source. For each OC, we estimate a suitable cut-off membership probability $\pcluster$ based on the $\pmem$ distribution of that OC and identify sources with $\pmem>\pcluster$ as cluster members \autoref{S:final_membership}.

Using these members, we estimate cluster global properties using isochrone fitting in a Bayesian framework without any need for human intervention (\autoref{S:cluster_props}). We identify the single isochrone from the available data and find representative polynomials depicting it (\autoref{S:cluster_props_MS}).  
We compare this with the theoretical isochrones generated using the \mist\ \citep{mist0, mist1} package. We use the \emcee\ package \citep{emcee} to estimate the posterior distributions for the cluster properties $\age$, $\metal$, $D$, and $\reddening$ by treating them as free parameters with appropriate priors. We summarize our key results below.

\begin{itemize}
\item For 19 clusters out of 21, we identify more cluster members than \cgmem. Our estimated member numbers are also higher than those found in \hunmem\ for all 21 OCs we have analyzed. The most likely explanation for this is that these studies may have preferentially missed low-mass MS binaries and stragglers. The increase in identified cluster members can be as high as a factor of $\approx2.9$ (\autoref{s:results_ncl}, \autoref{tab:cluster_properties_table}, \autoref{fig:all_violin_plot}).
\item We find the posterior distributions for important cluster properties by simultaneous Bayesian parameter estimation for $\age$, $\metal$, $D$, and $\reddening$ (\autoref{fig:2287_corner_main_plot}). In most cases our estimates agree well with those in relevant previous studies (\autoref{s:results_properties}, \autoref{fig:all_violin_plot}). In some cases we find some differences with likely reasons. For example, for some old OCs our estimated $\metal$ is lower compared to those in \citet{netopil_2016}, but in these cases, the estimates in \citet{netopil_2016} were based on very limited number of sources (\autoref{s:results_properties}).  
\item In order to make sure that we do not reject straggler members of an OC, we intentionally do not consider proximity of a source to any well-defined sequence on a CMD. Despite this, our identified members produce excellent clean CMDs. This bolsters our confidence that the members we identify are indeed members of the OCs. Moreover, our method is well-suited to identify straggler members which are important tracers of an OC's dynamical properties. 
\item Our isochrone fitting procedure does not require any human intervention. Moreover, although we only consider up to the MSTO for isochrone fitting, wherever there is a giant branch, our fitted isochrones simultaneously fits both the MS and the giant branch.  
\end{itemize}

In conclusion, we identify members of $21$ rich OCs using a non-parametric data-driven approach which takes into account correlated errors in astrometric and photometric errors for each source. We identify higher number of members compared to previous studies. Although in this study where we develop these methods, we have restricted ourselves to $\rproj/\pc=20$, we could easily extend the analysis to much higher $\rproj$. The only limitation is computational cost. Although several OC catalogs exist, often the properties are collected from a variety of studies employing disparate techniques (see discussion in \autoref{s:results_properties}). In this study we identify OC members and estimate property posterior distributions using methods that requires little subjectivity. Although, we employ our methods to 21 rich and nearby OCs in this first effort, we plan to expand this to more OCs in the future.  

\vspace{0.5cm}
 AG acknowledges support from TIFR’s graduate fellowship. PKN acknowledges TIFR's postdoctoral fellowship. PKN also acknowledges support from the Centro de Astrofisica y Tecnologias Afines (CATA) fellowship via grant Agencia Nacional de Investigacion y Desarrollo (ANID), BASAL FB210003. SC acknowledges support from the Department of Atomic Energy, Government of India, under project no. 12-R\&D-TFR-5.02-0200 and RTI 4002. All simulations are done using Azure cloud computing and TIFR-HPC.

\vspace{0.2cm}
\textit{Software} : For this work, we extensively use many open software projects like Python 3 \citep{python3}, Numpy \citep{numpy_scipy_matp}, scipy \citep{numpy_scipy_matp}, matplotlib \citep{numpy_scipy_matp}, pandas \citep{numpy_scipy_matp} and Astropy \citep{astropy:2013, astropy:2018}

\bibliography{main}
\bibliographystyle{aasjournal}

\appendix

\section{Surface denisty profile}\label{S:surface_density_profile}
We estimate the surface density profile of cluster members and all stars (field and cluster) in the vicinity of the cluster. In \autoref{fig:surface_deinsty_2287}, red (blue) dots show the cluster (overall) surface density profiles. The black horizontal solid line shows the average density of all stars in the 15-20 pc region of NGC-2287. The grey-shaded region is the corresponding error bar. From this, we notice the cluster surface density profile falls to a much lower value compared to that for all stars in the vicinity well below 20 pc. 

\noprint{\figsetstart}
\noprint{\figsetnum{14}}
\noprint{\figsettitle{Surface density profiles.}}

\figsetgrpstart
\figsetgrpnum{14.11}
\figsetgrptitle{NGC_2287}
\figsetplot{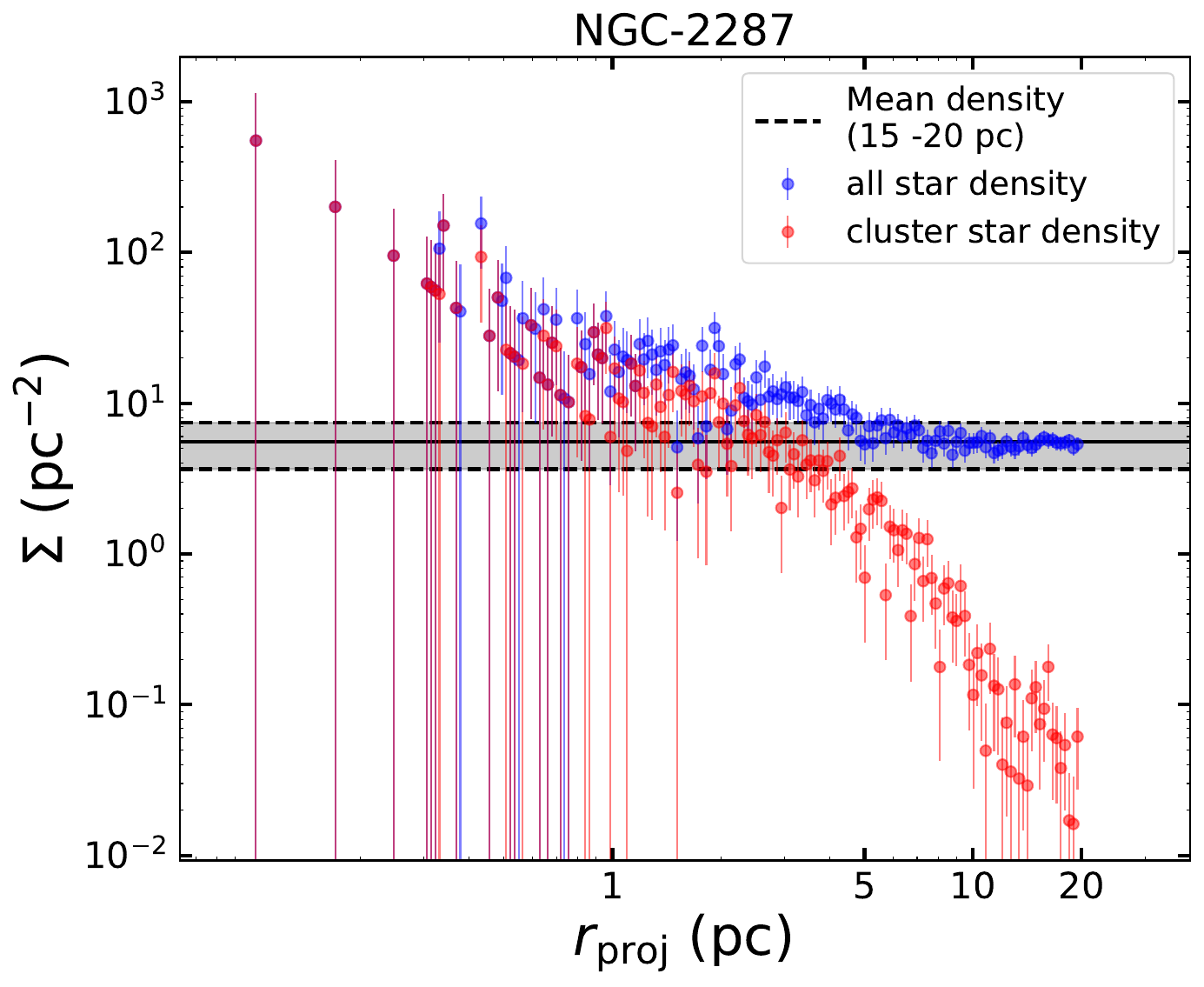}
\figsetgrpnote{Surface density profile for NGC_2287.}
\figsetgrpend

\figsetend

\begin{figure}[h]
\gridline{\fig{NGC_2287_surf_all_cluster_star.pdf}{0.85\textwidth}{}}
\caption{Surface density ($\Sigma$) profile for example cluster NGC-2287 (red). We also show the surface density profile for all sources, cluster members and field stars (red). The black solid line and the grey-shaded region show the mean $\Sigma$ for all sources within $15\leq\rproj/\pc\leq20$ and the corresponding $1\sigma$. We find that the $\Sigma$ for cluster members becomes much smaller than that for all sources at $\rproj$ significantly smaller than $20\,\pc$. This behavior is found in all 21 OCs we have analyzed.
}
\label{fig:surface_deinsty_2287}
\end{figure}
\section{Convergence Test}
\label{S:App-convergence_bayesian}
\begin{figure*}\label{convergence_corner_2287}
\gridline{\fig{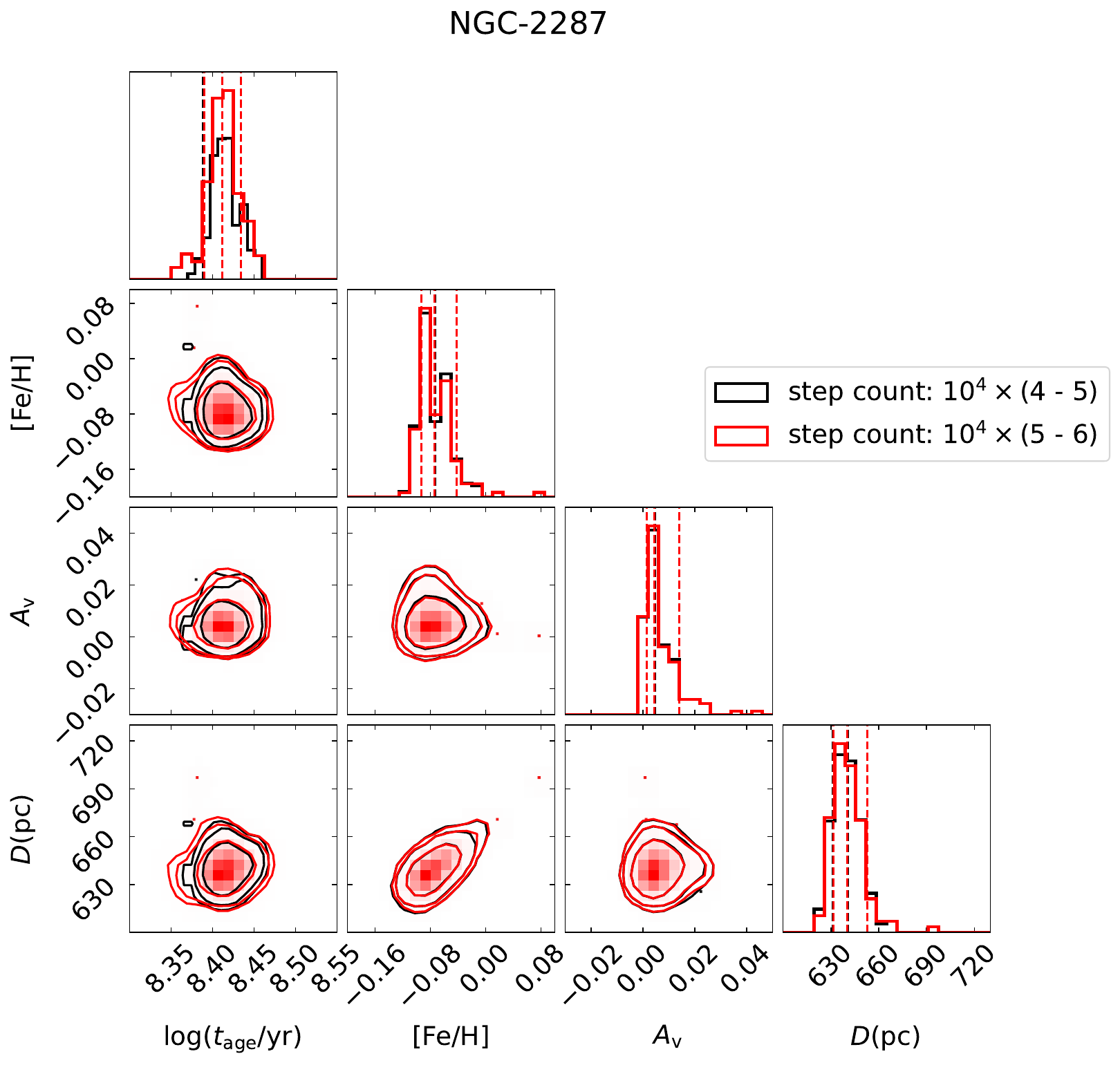}
{0.85\textwidth}{}}
\caption{Same as \autoref{fig:2287_corner_main_plot} but showing convergence of the posterior distributions at different stages of Bayesian parameter estimation for our example cluster NGC-2287. Black (red) shows the posterior distributions using walker positions between $40,000$--$50,000$ ($50,000$--$60,000$) step counts. Clearly the distributions show excellent agreement indicating that our MCMC chains have sufficiently converged.
}
\label{fig:2287_corner_covergence}
\end{figure*}

We use MCMC to evaluate the posterior distributions for clusters properties (\autoref{S:cluster_props_params}). For each cluster we use $128$ walkers, $60,000$ steps using the Metropolis-Hastings algorithm. For each case, we discard the first $30,000$ steps as burn in. We trim using 1 in 15 positions of each walker. Here we illustrate the convergence of the posterior distributions using our example cluster NGC-2287 (\autoref{fig:2287_corner_covergence}). We create two batches by collecting walker positions between $40,000$--$50,000$ steps and $50,000$--$60,000$ steps. We find that the posterior distributions created by these two batches are identical indicating that the number of steps we adopt is more than adequate and the posteriors have converged.

\section{Dependency on membership}
\label{S:App-pmemc}
\begin{figure}[h]
\gridline{\fig{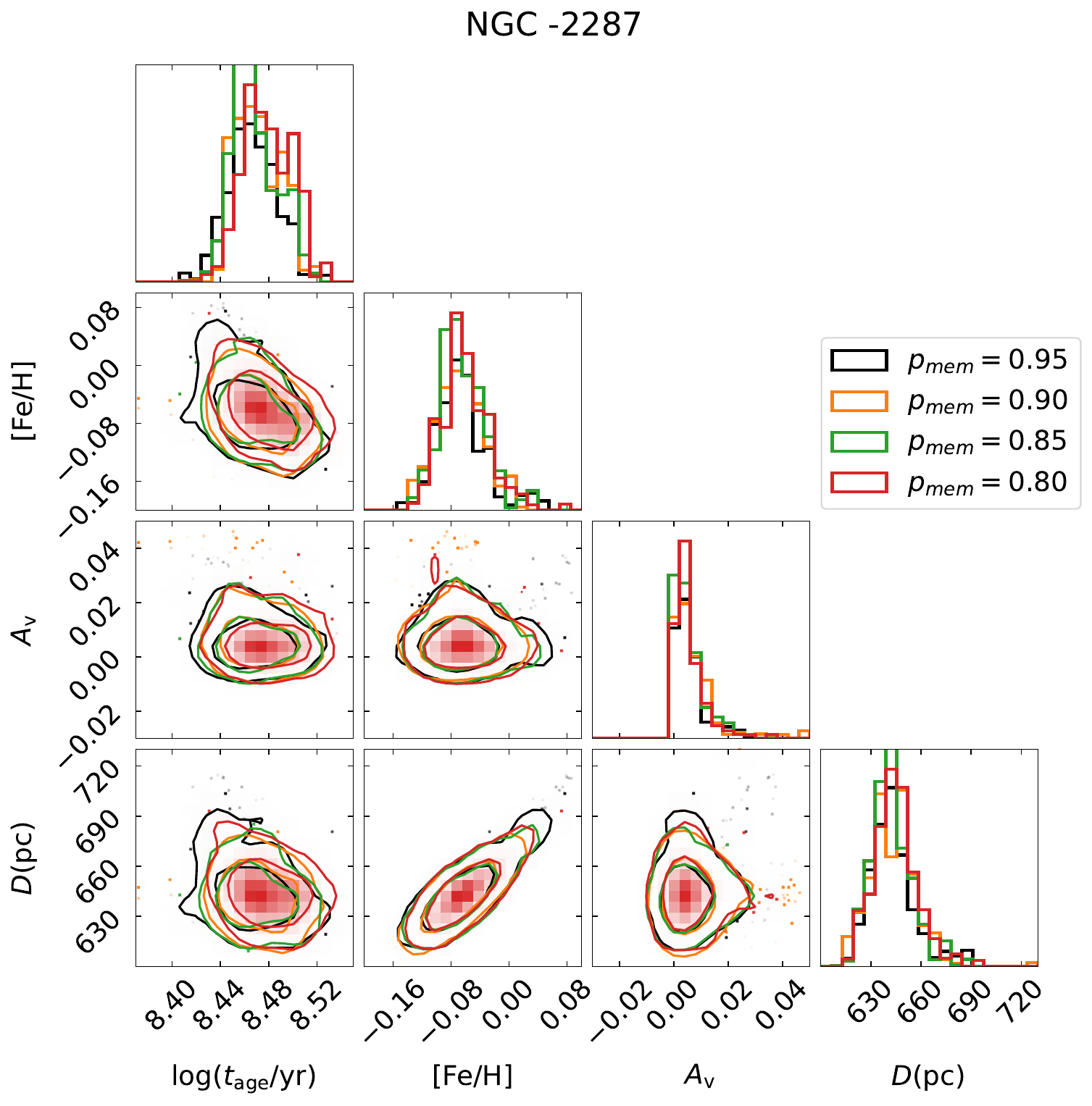}{0.85\textwidth}{}}
\caption{Same as \autoref{fig:2287_corner_main_plot} but showing the effect of adopted $\pcluster$ on the estimated cluster properties $\age$, $\metal$, $\reddening$, and $D$. The different colors denote different adopted $\pcluster$ (see legend). We find that the adopted $\pcluster$ (within reason) does not affect the final posterior distributions of cluster properties.}
\label{fig:comparison_mem_iso}
\end{figure}

The choice of cut-off membership probability adopted in our study, $\pcluster$, has been a matter of choice. Here we investigate how the cluster global properties may change with different adopted $\pcluster$ values. We collect the cluster members for different cluster membership values and use them in our Bayesian framework to estimate cluster properties. We show in \autoref{fig:comparison_mem_iso} that the final posterior distribution of cluster properties is almost the same for different adopted $\pcluster$. 

\end{document}